\documentclass{elsarticle}

\usepackage[utf8]{inputenc}

\usepackage{amsmath,amssymb,amsfonts}
\usepackage{graphicx}
\usepackage{hyperref}
\usepackage{url}

\usepackage{fancyvrb}
\usepackage{xspace}
\usepackage{mdframed}

\usepackage{colortbl}
\usepackage{xcolor}
\usepackage{multirow}
\usepackage{caption}

\newcommand{\caret}{$^{\wedge}$}

\newcommand{\etal}{et al.\xspace}
\newcommand{\ie}{i.e.,\xspace}
\newcommand{\eg}{e.g.,\xspace}
\newcommand{\fig}[1]{Figure~\ref{#1}}
\newcommand{\tab}[1]{Table~\ref{#1}}
\newcommand{\sect}[1]{Section~\ref{#1}}

\newcommand{\semver}{\textsf{semver}\xspace}

\newcommand{\preone}{{0.y.z}\xspace}
\newcommand{\postone}{{$\geq$1.0.0}\xspace}

\newcommand{\cargo}{\textsf{Cargo}\xspace}
\newcommand{\npm}{\textsf{npm}\xspace}
\newcommand{\packagist}{\textsf{Packagist}\xspace}
\newcommand{\rubygems}{\textsf{RubyGems}\xspace}
\newcommand{\haskell}{\textsf{Haskell}\xspace}

\newcommand{\github}{GitHub\xspace}

\newcommand{\javascript}{\textsf{JavaScript}\xspace}
\newcommand{\php}{\textsf{PHP}\xspace}
\newcommand{\rust}{\textsf{Rust}\xspace}
\newcommand{\ruby}{\textsf{Ruby}\xspace}

\newcommand{\figsize}{0.9\columnwidth}

\newcommand{\changed}[1]{\xspace{#1}\xspace}

\newenvironment{custombox}{\medskip\begin{mdframed}[topline=false, bottomline=false, rightline=false, linewidth=12pt, linecolor=lightgray!80, backgroundcolor=lightgray!20, roundcorner=2pt, innerleftmargin=12pt, innerrightmargin=12pt, innertopmargin=12pt, innerbottommargin=12pt, nobreak=true]}{\end{mdframed}\medskip}

\journal{Science of Computer Programming}

\begin{document}

\begin{frontmatter}
\title{Lost in Zero Space -- An Empirical Comparison of\\ \changed{0.y.z} Releases in Software Package Distributions}

\author[1]{Alexandre Decan\corref{cor1}}
\ead{alexandre.decan@umons.ac.be}
\author[1]{Tom Mens}
\ead{tom.mens@umons.ac.be}

\address[1]{Software Engineering Lab, University of Mons, Avenue Maistriau 15, Mons 7000, Belgium}
\cortext[cor1]{Corresponding author}

\begin{abstract}
Distributions of open source software packages dedicated to specific programming languages facilitate software development by allowing software projects to depend on the functionality provided by such reusable packages.
The health of a software project can be affected by the maturity of the packages on which it depends.
The version numbers of the used package releases provide an indication of their maturity.
Packages with a \preone version number are commonly assumed to be under initial development, suggesting that they are likely to be less stable, and depending on them may be \changed{considered as} less healthy.

In this paper, we empirically study, for four open source package distributions (\cargo, \npm, \packagist and \rubygems) to which extent \preone package releases and \postone package releases behave differently. We quantify the prevalence of \preone releases, we explore how long packages remain in the initial development stage, we compare the update frequency of \preone and \postone package releases, we study how often \preone releases are required by other packages, we assess whether semantic versioning is respected for dependencies towards them, and we compare some characteristics of \preone and \postone package repositories hosted on \github.
Among others, we observe that package distributions are more permissive than what semantic versioning dictates for \preone releases, and that many of the \preone releases can actually be regarded as mature packages. 
As a consequence, the version number does not provide a good indication of the maturity of a package release.
\end{abstract}


\end{frontmatter}




\section{Introduction}
\label{sec:intro}

Open source software development embraces the principles of software reuse, through the availability of software package distributions dedicated to specific programming languages (\eg \cargo for \rust, \npm for \javascript, \rubygems for \ruby, and \packagist for \php).
As is the case for any software system, the reusable packages in such distributions can have different levels of maturity.
In order for a mature software project to be considered healthy, it should avoid depending on unstable and immature reusable packages that are still in their initial development phase.

A common convention for packages to reflect this maturity in their version number \textsf{major.minor.patch} is to set the major version component to \textsf{1} as soon as they reach their first stable release: \emph{``in open source, tagging the 1.0.0 release is comparable to shipping a final product''}~\cite{JeremyKahnBlog2013}, \emph{``1.0.0 indicates some degree of production readiness''}~\cite{RedditFear100}.
Packages that are under initial development assign a major version number \textsf{0} to their releases, conveying that the software is still incomplete and remains work in progress.
A \preone version number can therefore be seen as a signal to treat the package differently than a \postone version. \changed{A JavaScript developer states this as follows:} \emph{``it means that a project cannot be trusted. It would be unwise for a business-critical application to have a dependency that is young and prone to significant changes at any time. It also indicates that the dependency project simply isn’t done and might not be ready for use.''}~\cite{JeremyKahnBlog2013}.

If we assume that this convention is followed, it would be advisable for \postone packages to not depend on such \preone packages. \changed{Similarly, it would be advisable} for \preone packages to quickly reach a \postone release in order to allow other packages to depend on a stable and mature release.
\changed{As witnessed by a \rust developer, \emph{``some reasonably mature crates are still at version \preone or depend on other crates that are at \preone. This is seen as a bad thing and usually results in a few \github issues pushing the authors of those crates to change the version to 1.0.''}~\cite{RedditMeaning100}}
In practice, however, many packages remain \emph{stuck in zero version space}.
\changed{There seems to be a psychological barrier associated to crossing the 1.0.0 version} that package maintainers may associate with additional responsibilities.

Some versioning policies explicitly materialise the differences between \preone and \postone versions.
Consider for example \semver (semantic versioning)~\cite{semver2}, a common versioning policy in package distributions~\cite{Bogart2017survey}, dictating how version numbers should be incremented w.r.t. backward compatibility.
This policy distinguishes \preone from \postone versions in terms of maturity, release cycle and stability.
\changed{For example, \semver enables to assess the backwards compatibility of a \postone release based on its version number by distinguishing between incompatible changes (\ie an increment in the major component of the version number) and compatible changes (\ie an increment in the minor or patch component).
For \preone releases on the other hand, \semver considers that \emph{``anything may change at any time''}. This could be problematic for developers of dependent packages, since they need to find other ways to assess the compatibility of such releases. They could also decide to stay on the safe side by preventing their installation, implying they will not benefit from the bug and security fixes provided by the new version.}

The specific rules for \preone versions are sometimes considered disruptive\footnote{\url{https://github.com/semver/semver/issues/221}} and counter-intuitive.\footnote{\url{https://github.com/npm/node-semver/issues/79}}
The term \emph{magic zero} reflects this different semantics for \preone versions.
The confusion around magic zero notably led the maintainers of \npm to recommend package developers to avoid using \preone version numbers and start from version 1.0.0 ``\emph{since the \semver spec is weirdly magical about 0.x.y versions, and we cannot ever hope to get everyone to believe what the correct interpretation of 0.x versions are.}''\footnote{\url{https://github.com/npm/init-package-json/commit/363a17bc31bf653}}
This also gave rise to the satirical \emph{ZeroVer} versioning policy, stating that ``\emph{your software's major version should never exceed the first and most important number in computing: zero}''.\footnote{\url{https://0ver.org}}

The goal of this article is to assess quantitatively to what extent package developers in different package distributions take into account such differences in package releases. 
To reach this goal, we study the following research questions in  \cargo, \npm, \packagist and \rubygems, four package distributions that are known to adhere to \semver~\cite{decan2019tse}:

\begin{description}
    \item[$RQ_1$:] {\em How prevalent are \preone packages?}
    We observe in all distributions that a high proportion of packages did not yet reach a \postone release.

    \item[$RQ_2$:] {\em Do packages get stuck in the zero version space?}
    The overwhelming majority of packages never traversed the 1.0.0 barrier,
    and of those that did, more than one out of five took more than a year to do so.

    \item[$RQ_3$:] {\em Are \preone \changed{releases published} more frequently than \postone \changed{releases}?}
    Although a statistical difference could be observed, this difference was small to negligible in each distribution.

    \item[$RQ_4$:] {\em Are \preone package releases required by other packages?}
    This was indeed observed as a frequent phenomenon for each package distribution.

    \item[$RQ_5$:] {\em How permissive are the dependency constraints towards required \preone packages?}
    Most dependency constraints towards \preone packages accept new patches, making them more permissive than what \semver recommends. According to \semver, \emph{``Major version zero (0.y.z) is for initial development. Anything may change at any time. The public API should not be considered stable''}~\cite{semver2}.

    \item[$RQ_6$:] {\em Do \github repositories for \preone packages have different characteristics?}
    \github repositories associated to \preone packages were found to have slightly less stars, forks, contributors\changed{, open issues} and dependent repositories, and to be slightly smaller than repositories for \postone packages.

\end{description}

This paper builds further upon previous work~\cite{Decan2020Zero} in which we empirically analysed the prevalence of \preone releases in \cargo, \npm and \packagist, and observed some preliminary evidences of the (lack of) differences between \preone and \postone packages.
The current paper extends this study by including a fourth package distribution (namely \rubygems) and one year of extra data for the other three distributions, accounting for more than 500K additional packages for the empirical study.
We carry out deeper analyses to complement the preliminary insights we got, such as studying the activity of \preone and \postone packages ($RQ_1$), the time it takes to cross the magic 1.0.0 barrier ($RQ_2$), the \changed{evolution of the release frequency} of \preone and \postone releases ($RQ_3$) and the number of dependent packages ($RQ_4$) for \preone and \postone packages, and the evolution of dependency constraints in \preone releases ($RQ_5$).
We also added an entirely novel research question on the characteristics of \preone and \postone package repositories on \github ($RQ_6$)\changed{. We} complement our discussions with anecdotal evidence from developers, and an analysis of the version number\changed{s} used for initial package releases.

The remainder of this paper is structured as follows.
\sect{sec:related} discusses related work.
\sect{sec:methodology} introduces the research methodology.
\sect{sec:rqs} empirically studies the research questions and presents the main findings.
\sect{sec:discussion} discusses the results and
\sect{sec:threats} presents the threats to validity of the research, and \sect{sec:conclusion} concludes.

\section{Related Work}
\label{sec:related}


The staged software life cycle model~\cite{BennettRajlich1999} suggests that the \emph{initial development} stage of software should be distinguished from its subsequent evolution: ``\emph{During initial development, engineers build the first functioning version of the software from scratch to satisfy initial requirements.}''
This staged model has been studied in the context of open source software projects by Capiluppi \etal~\cite{capiluppi2007adapting}, who observed that ``\emph{many FLOSS projects could be argued to never have left this [initial development] stage}''.
Fernandez \etal~\cite{fernandez2008empirical} observed that the software evolution laws apply well to open source projects having achieved maturity. They confirm, however, that ``\emph{many projects do not pass the initial development stage.}''
Similarly, Costa \etal~\cite{costa2018sustainability} observed that 44 out of 60 evolving academic software projects
(\ie 73\%) are either in initial development or closedown stage.


The typical way to distinguish initial software development releases from stable ones is by resorting to some kind of versioning scheme.
For software libraries, the most common approach appears to be to use some variant of \textsf{major.minor.patch} version numbers. The expressiveness of such a version numbering has been accused of being too limited~\cite{stuckenholz2005component}:
``\emph{Especially if component developers need to assign version numbers to their components manually and do not have proper instructions that define which changes in what level of contract conduct a new version, those version numbers at most rest for marketing use and do not ensure compatibility between different components.}''
%
%
The \semver policy~\cite{semver2} was introduced in an attempt to address such issues, and to provide a partial solution to the ``dependency hell'' that developers face when reusing software packages.
It conveys a meaning to the \textsf{major.minor.patch} version number, assuming that the reusable package has a public API: ``\emph{Bug fixes not affecting the API increment the patch version, backwards compatible API additions/changes increment the minor version, and backwards incompatible API changes increment the major version.}''

The use of \semver is quite common for package distributions~\cite{Bogart2017survey}.
Wittern \etal~\cite{Wittern2016} studied the evolution of a subset of \npm packages, analysing characteristics such as their dependencies, update frequency, popularity, and version numbering.
They found that package maintainers adopt numbering schemes that may not fully adhere to the semantic versioning principle; and that a large number of package maintainers are reluctant to ever release a version 1.0.0.
Raemaekers \etal~\cite{Raemaekers2017JSS} investigated the \semver-compliance in 22K \textsf{Java} libraries in \textsf{Maven} over a seven-year time period.
They found that breaking changes appear in one third of all releases, including minor releases and patches, implying that \semver is not a common practice in \textsf{Maven}.
Because of this, many packages use strict dependency constraints and package maintainers avoid upgrading to newer versions of dependent packages.
Decan \etal~\cite{Decan2017SANER}  studied the use of package dependency constraints in \npm, \textsf{CRAN} and \rubygems.
They observed that, while strict dependency constraints prevent backward incompatibility issues, they also increase the risk of having dependency conflicts, outdated dependencies and missing important updates.
Decan \etal~\cite{decan2019tse} studied \semver-compliance in four evolving package distributions (\cargo, \npm, \packagist and \rubygems).
They observed that these distributions are becoming more \semver-compliant over time, and that \changed{package distributions use} specific notations, characteristics, maturity and \changed{policies that} play an important role in the degree of such compliance.


Bogart \etal~\cite{Bogart2016} qualitatively compared \npm, \textsf{CRAN} and \textsf{Eclipse}, to understand the impact of community values, tools and policies on breaking changes. They identified two main types of mitigation strategies to reduce the exposure to changes in dependencies: limiting the number of dependencies, and depending only on ``trusted packages''.
They also found that policies and practices may diverge when policies are perceived to be misaligned with the community values and the platform mechanisms.
They confirmed this in a follow-up qualitative study about values and practices in 18 software package distributions, on the basis of a survey involving more than 2,000 developers~\cite{Bogart2017survey}.
Different package distributions were found to have different priorities and make different value trade-offs.
Their results show that relationships between values and practices are not always straightforward.

\section{Data Extraction}
\label{sec:methodology}
\label{sec:data}

In previous work we studied the use of \semver in four package distributions (\cargo, \npm, \packagist and \rubygems)~\cite{decan2019tse} and observed that these distributions where mostly \semver-compliant. We also observed intriguing differences between how pre-1.0.0 and post-1.0.0 dependency constraints were being used. This triggered us to conduct the current in-depth study on the presence and use of \preone package releases in these package distributions.

To analyze the considered package distributions, we rely on version 1.6.0 of \textsf{libraries.io} Open Source Repository and Dependency Metadata~\cite{Katz2020}, released in January 2020.
\changed{This dataset contains, among others, the metadata of packages in \cargo, \npm, \packagist and \rubygems. These metadata include all package releases, their version number, their release date, their dependencies including the target package, the dependency constraint and the scope of the dependency (\eg runtime, test, ...). The dataset also contains various metadata for the git repositories related to these packages, such as the address of the repository, the number of contributors, the number of forks, stars, etc.}

For each package distribution, we consider all packages and all their releases, except for the pre-release versions (such as 2.1.3-alpha, 0.5.0-beta or 3.0.0-rc) that are known to be ``\emph{unstable and might not satisfy the intended compatibility requirements as denoted by its associated normal version}''~\cite{semver2}.
For each package release, we consider only dependencies to other packages within the same distribution, \ie we ignore dependencies targeting external sources (\eg websites or git repositories).
%
\changed{When declaring dependencies, a package maintainer can specify the purpose of the dependency (\eg it is needed to execute, develop or test the package). We excluded dependencies that are only needed to test or develop a package because not all considered packages make use of them, and not every package declares a complete and reliable list of such dependencies. We therefore consider only those dependencies that are required to install and execute the package, and hence more accurately reflect what is needed to actually use the package.}
%
In the package distributions we analyzed, \changed{these dependencies are either marked as} ``runtime'' or ``normal''.


To reduce noise in the dataset, we \changed{manually inspected and removed outlier packages} with clearly deviating and undesirable behaviour.
For \packagist we excluded 23 of the most active packages (and their associated 1.2K releases) that were created and published to promote illegal download services.
For \npm we excluded around 23K packages (and their associated 51K releases) that were purposefully created by malevolent developers abusing the API of the \npm package manager.
These are either packages whose main purpose is to depend on a very large number of other packages (\eg \textsf{npm-gen-all}) 
or replications and variations of existing packages (\eg \textsf{npmdoc-*}, \textsf{npmtest-*}, \textsf{*-cdn}, etc.)
Most of them are no longer available through \npm.

\begin{table}[!h]
    \caption{Characteristics of the curated dataset.}
    \label{tab:dataset}
    \centering

    \begin{tabular}{c|ccrrr}
        \bf distribution & \bf created & \bf language & \bf \#pkg & \bf \#rel & \bf \#dep\\
        \hline\hline
         \cargo & 2014 & \rust & 35K & 183K & 796K\\
         \npm & 2010 & \javascript & 1,218K & 9,383K & 48,695K\\
         \packagist & 2012 & \php & 180K & 1.520K & 4,727K\\
         \rubygems & 2004 & \ruby & 155K & 956K & 2,396K\\
        \hline
        \multicolumn{3}{l}{} & 1,588K & 12,041K & 56,615K\\
    \end{tabular}

\end{table}

\tab{tab:dataset} summarises the curated dataset, reporting the number of packages (\#pkg), package releases (\#rel), and dependencies (\#dep) that are considered for the empirical analysis.
The data and code to replicate the analysis are available on \url{https://doi.org/10.5281/zenodo.4013419}.

\section{Research Questions}
\label{sec:rqs}

\subsection{How Prevalent Are \preone Packages?}
\label{sec:prevalence}

\begin{figure}[!htbp]
    \centering
    \includegraphics[width=\figsize]{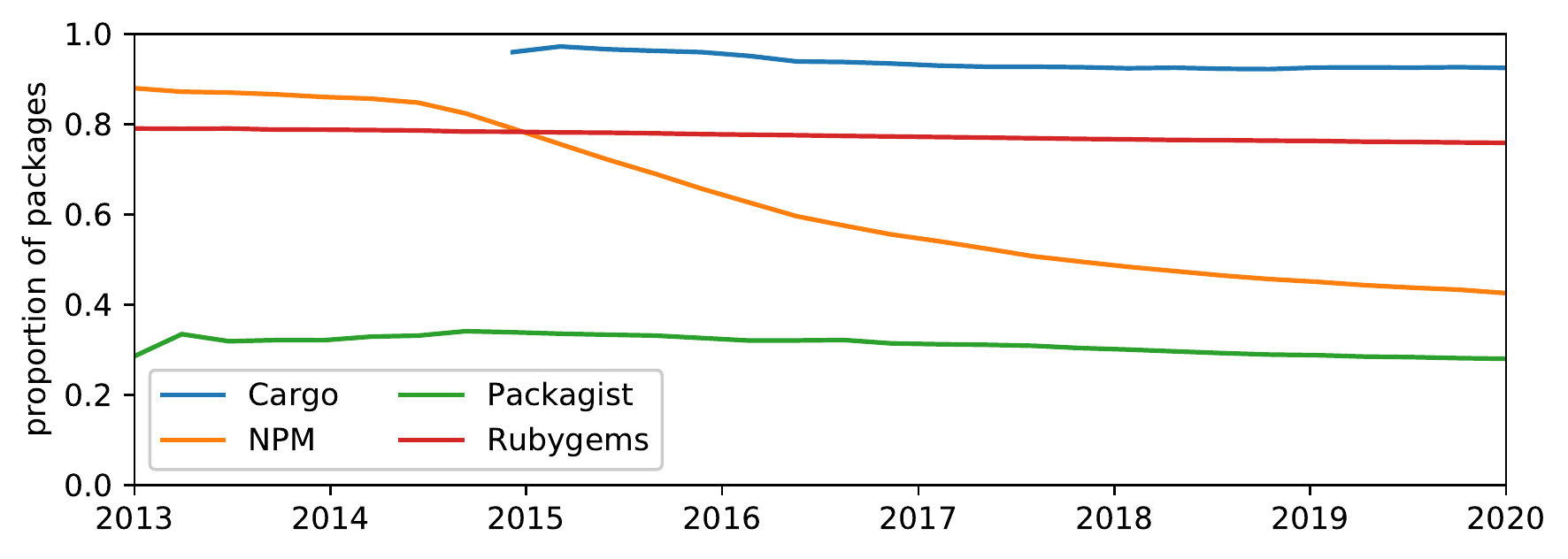}
    \caption{
        Evolution of the proportion of packages whose latest release is \preone.
    }
    \label{fig:prevalence_packages}
\end{figure}

Since the results of our analysis are only relevant if a sufficient number of packages in each distribution are still in their initial development phase, we compute for each distribution, on a monthly basis, the proportion of packages whose latest available release is \preone.
\fig{fig:prevalence_packages} shows the evolution of this proportion relative to the number of packages distributed at that time.

We observe a very high proportion of initial development packages for \cargo (94\% on average).
At the last observation time (December 2019), 92.4\% of its packages were still \preone packages.
For \rubygems, we observe a quite stable proportion of \preone packages over time, of 77.4\%.
On the other side of the spectrum we find \packagist, with 31\% of \preone packages on average, and 30\% at the last observation time.
\npm is the only one of the considered distributions to exhibit a non stable proportion of \preone packages.
Indeed, we observe from April 2014 onwards that the proportion of \preone packages went from 85.5\% to 42.5\%.
This is a consequence of \npm policies aiming to reduce the use of \preone releases, notably by changing the initial version of packages created through \texttt{npm init} to 1.0.0 instead of 0.1.0 \changed{``\emph{since the \semver spec is weirdly magical about 0.x.y versions, and we cannot ever hope to get everyone to believe what the correct interpretation of 0.x versions are.}''\footnote{See \url{https://github.com/npm/init-package-json/commit/363a17bc3}}}

\medskip

\begin{figure}[!htbp]
    \centering
    \includegraphics[width=\figsize]{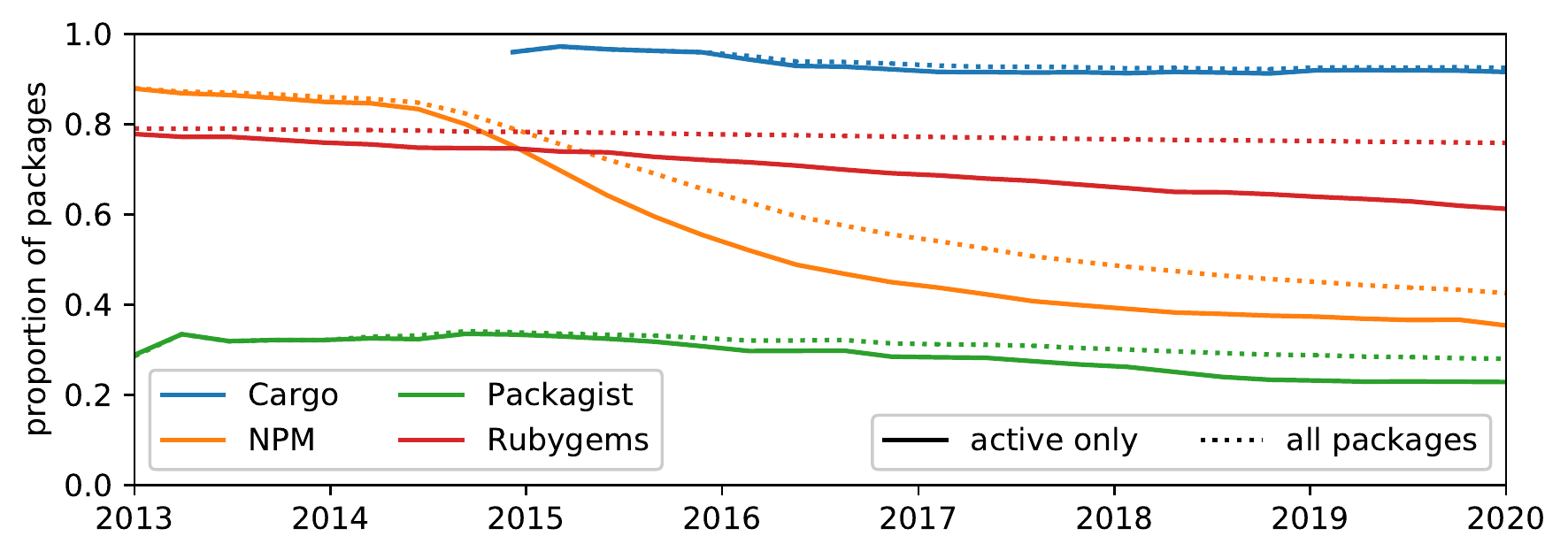}
    \caption{
        \changed{Evolution of the proportion of active \preone packages (straight lines).}
    }
    \label{fig:prevalence_packages_combined}
\end{figure}

It could be the case that the high proportions of \preone packages we observed are due \changed{to \emph{``many open source projects [that] languish in the 0.x.x state because the developer lost interest and stopped working on it''}~\cite{JeremyKahnBlog2013}}, \ie due to packages no longer being maintained and thus preventing them from ever reaching a \postone release.
To check whether inactive packages could have influenced the observations we \changed{derived} from \fig{fig:prevalence_packages}, we repeated this analysis by removing all packages that were not active \changed{during the last year (\ie packages that have not released a new version during the last 12 months).}
\changed{\fig{fig:prevalence_packages_combined} shows the evolution of the proportion of {\em active} \preone packages (straight lines) relative to the number of active packages. To ease the comparison, we also report the proportion of {\em all} (\ie both active and inactive) \preone packages (dotted lines) as in \fig{fig:prevalence_packages}.}
\footnote{\changed{The proportion of \postone packages can be obtained by taking the complement of the proportion of \preone packages.}}

\changed{We observe that the proportions of {\em active} \preone packages (straight lines) follow closely the proportions of {\em all} \preone packages (dotted lines).
This is especially visible in \cargo where the difference is only of $0.8\%$ ($=92.4 - 91.6$) in the last considered snapshot.
We also observe that the difference remains limited for for \packagist ($5.2\%=28 - 22.8$) and for \npm ($7.2\%=42.5 - 35.3$), indicating that the proportion of \preone packages does not depend on whether they are active or not.
Even if the difference is more pronounced in \rubygems ($14.6\% =75.8 - 61.2$), there are still proportionally many more \preone than \postone packages amongst the active ones.
We can therefore reject} our hypothesis that the high proportions of \preone packages \changed{we observed in \fig{fig:prevalence_packages} are} due to the presence of many \preone packages being no longer maintained.

\medskip

\begin{figure}[!htbp]
    \centering
    \includegraphics[width=\figsize]{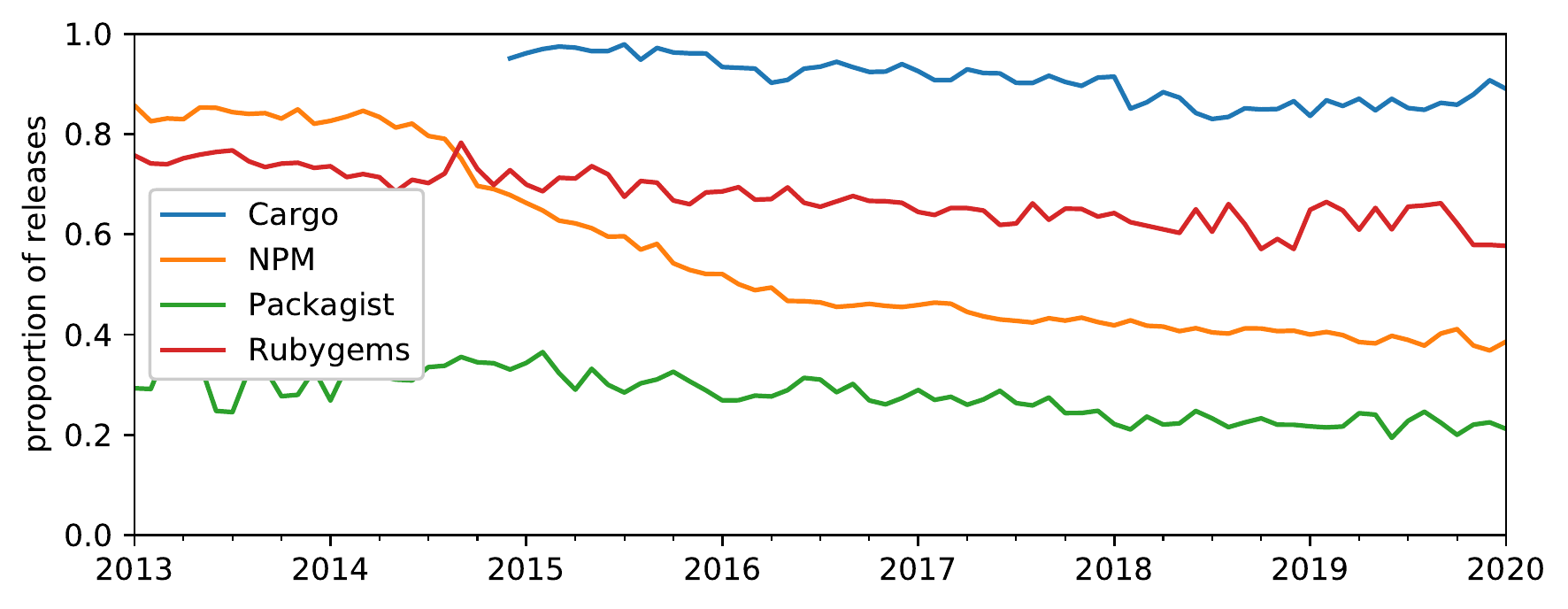}
    \caption{
        Evolution of the monthly proportion of \preone \changed{releases}.
    }
    \label{fig:prevalence_updates}
\end{figure}

To confirm that \preone packages represent a large part of the activity in the considered distributions, we computed the monthly proportion of \preone \changed{releases relatively to the total number of new releases}.
\fig{fig:prevalence_updates} shows the evolution of these proportions.
We observe that \preone packages are responsible for the large majority of package \changed{releases} in \cargo (median 90.8\%),  \rubygems (median 74.3\%) and \npm (median 58.8\%).
On the other hand, ``only'' one out of four \changed{releases} in \packagist is due to \preone packages (median 27.1\%).
We also observe that these proportions are slightly decreasing over time in all package distributions.
This is especially visible for \npm from April 2014 onwards, a consequence of the new \npm policies about initial development releases, as mentioned above.
However, at the end of the observation period, \preone packages still account for 89\% of all package \changed{releases} in \cargo, for 57.7\% in \rubygems, for 38.6\% in \npm and for 21.2\% in \packagist.
These proportions are quite similar to the proportions of active \preone packages in each distribution, indicating that \preone and \postone packages do not really differ in term of release activity.

\begin{custombox}
The considered package distributions contain many \preone packages:
more than one out of five in \packagist, more than one out of three in \npm, more than three out of five in \rubygems, and more than nine out of ten in \cargo.
\preone packages are nearly as active as \postone packages, and are responsible for
the majority of all package \changed{releases} in \cargo and \rubygems.\\
The release policies of \cargo and \rubygems should be adapted to incite package maintainers to move out of the zero version space, the same way as \npm has successfully done in 2014.
\end{custombox}

\subsection{Do Packages Get Stuck in the Zero Version Space?}
\label{sec:time}

If one assumes that \preone packages are still under initial development, then they are eventually expected to reach a \postone release reflecting their maturation.
However, developers are \emph{``hesitant to increment their projects to 1.0.0 and stay in 0.x.x for a very long time, and possibly forever.''}~\cite{JeremyKahnBlog2013}

\changed{To measure how long it takes to reach a \postone release in each of the considered package distribution}, we rely on the statistical technique of survival analysis (a.k.a. event history analysis)~\cite{Aalen2008} to model the time for the event ``package reaches \postone'' to occur as a function of the time elapsed since the first release of that package.
Survival analysis estimates the survival rate of a given population of subjects (packages in our case), \ie the expected time duration until the event of interest (\changed{from the first available release until the first} \postone release) occurs.
\changed{Survival analysis takes} into account the fact that some observed subjects may be ``censored'', either because the event was observed prior to the observation period (\ie the first considered release of the package was already \postone), or not observed during the observation period (\ie the package never reached a \postone release).
A common non-parametric statistic used to estimate survival function is the Kaplan-Meier estimator \cite{kaplan-meier}.

\begin{figure}[!htbp]
    \centering
    \includegraphics[width=\figsize]{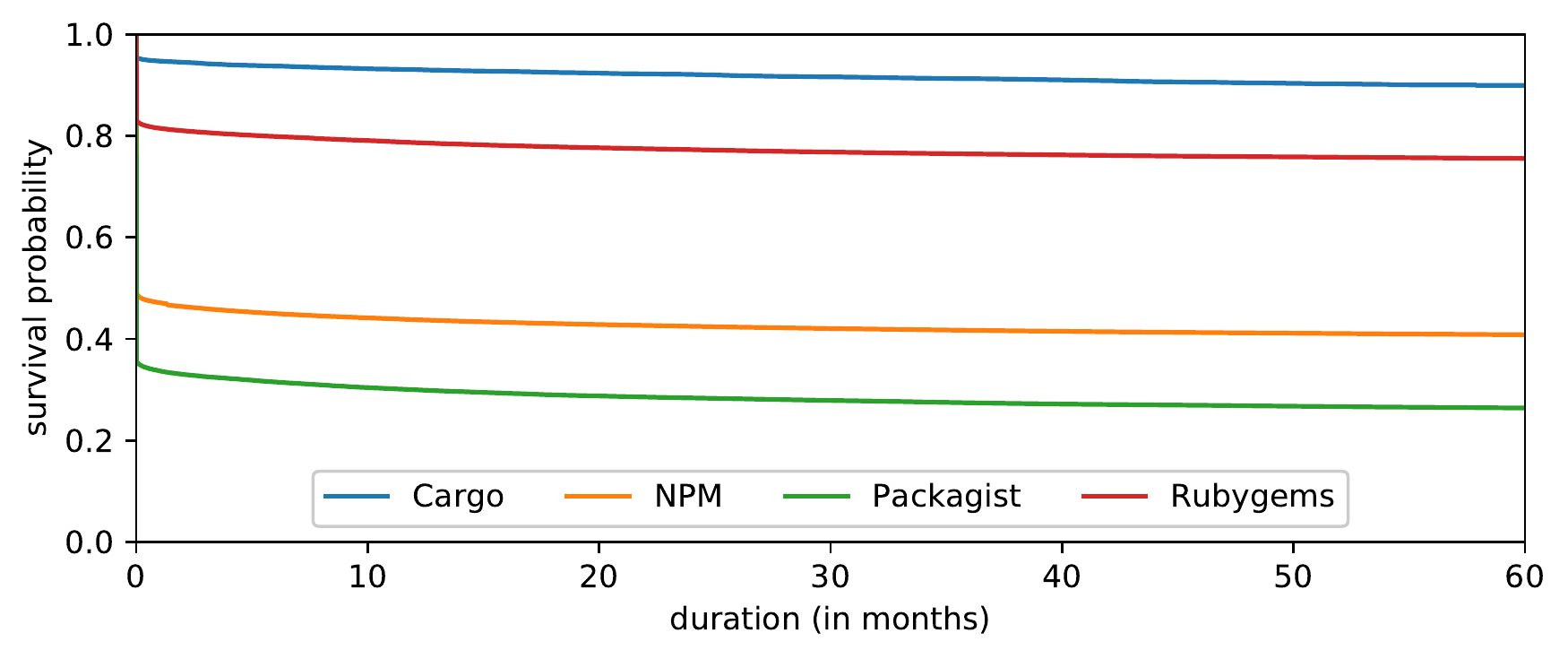}
    \caption{
        Kaplan-Meier survival curves for the probability that a ``package reaches \postone'' as a function of the time elapsed since the first release of that package.
   }
    \label{fig:barrier_uncensored}
\end{figure}

Survival functions define the probability of surviving past time $t$ or, equivalently, the probability that the event has not occurred yet at time $t$ (i.e., a package in the distribution has not reached a \postone release). \fig{fig:barrier_uncensored} shows the Kaplan-Meier survival functions for the four considered distributions.
Based on this probability, the complement probability that a package reaches \postone within a given time can be easily computed.

We observe that the survival probability over time is mainly driven by the survival probability at time 0, \ie by the proportion of packages being directly released with a \postone release.
While the probability to remain \preone slightly decrease over time, this decrease is rather limited (from 0.06 for \cargo to 0.10 for \packagist).
On average, the survival probability decreases by less than 0.02 per year, a direct consequence of many packages having not reached yet the 1.0.0 barrier.

\medskip

\begin{figure}[!htbp]
    \centering
    \includegraphics[width=\figsize]{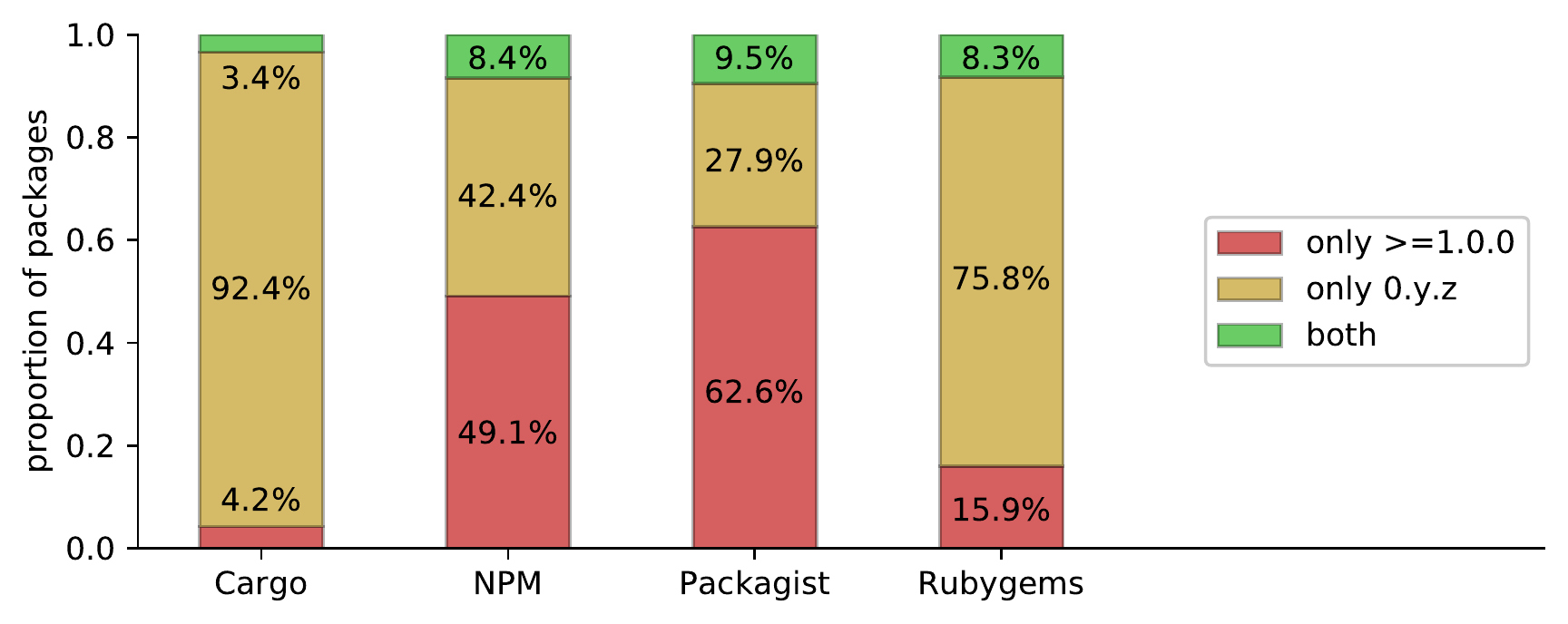}
    \caption{
        Proportion of packages having only \postone releases, only \preone releases, or both.
    }
    \label{fig:barrier_numbers}
\end{figure}

To confirm \changed{that most packages have not reached the 1.0.0 barrier yet}, we computed the proportion of packages whose first distributed release was already mature (\postone), packages that eventually crossed the 1.0.0 barrier, 
and packages remaining in their \preone phase.
\fig{fig:barrier_numbers} shows the proportion of such packages for each package distribution.

We observe that most packages in \cargo (92.4\%) and in \rubygems (75.8\%) only have \preone releases.
On the other hand, the majority of \packagist packages (62.6\%) only have \postone releases.
For \npm, there is a more or less equal proportion of packages having only \preone releases and packages having only \postone releases.
We also observe that, regardless of the package distribution, less than one out of ten packages went from a \preone to a \postone release.
This represents 1.176 packages in \cargo (3.4\%), around 103K in \npm (8.4\%), 17K packages in \packagist (9.5\%) and 13K packages in \rubygems (8.3\%).

\medskip

\begin{figure}[!htbp]
    \centering
    \includegraphics[width=\figsize]{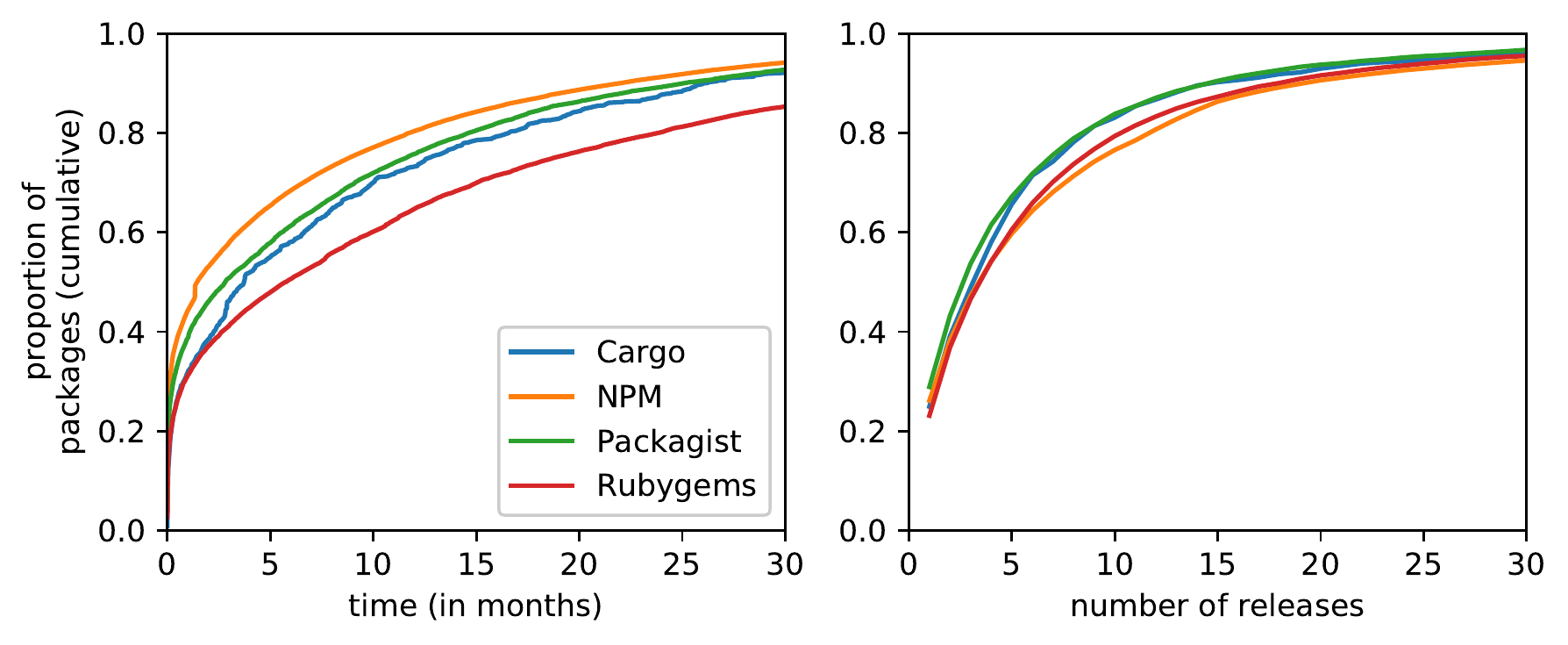}
    \caption{
        Duration (in months) and number of releases between first \preone and first \postone release.
    }
    \label{fig:barrier_combined}
\end{figure}

Focusing exclusively on packages that traversed the 1.0.0 barrier, we computed the duration between their first \preone release and their first \postone release.
\fig{fig:barrier_combined} presents the cumulative proportion of packages having reached the 1.0.0 barrier in function of the duration in time (left) and in terms of number of intermediate releases (right).

We observe that a majority of packages take only a few months and a few release updates to reach \postone. The median duration varies between 1.4 months (for \npm) and 5.8 months (for \rubygems) while the median number of release updates is 3 for \packagist and 4 for the other package distributions.
Yet there are many packages that took much longer to reach \postone.
Over one out of five packages (35.3\% for \rubygems, 26.7\% for \cargo, 24.3\% for \packagist and 19.6\% for \npm) needed more than a year to reach \postone, and many packages more than 2 years (19.9\% in \rubygems, 12.3\% in \cargo, 10.8\% in \packagist and 8.8\% in \npm).
There are even 6.2\% of all packages in \rubygems that needed more than 4 years to reach \postone. For comparison, they are less than 1.8\% in the other distributions.

\begin{custombox}
Many packages get stuck in the zero version space.
The probability to reach \postone increases by less than 0.02 per year.
Less than 10\% of all packages went from \preone to \postone releases.
While a majority of them only took a few months and a few updates to reach \postone, one out of five of them took more than one year to cross the 1.0.0 barrier, and many of them took even more than two years, especially in \rubygems.\\
Package maintainers should not be afraid of crossing the 1.0.0 barrier. Packages that have been developed for years and that are indubitably ready for production should receive a major version 1 or higher.
\end{custombox}

\subsection{Are \preone \changed{Releases Published} More Frequently \changed{than \postone releases}?}
\label{sec:freq}

\begin{figure}[!htbp]
    \centering
    \includegraphics[width=\figsize]{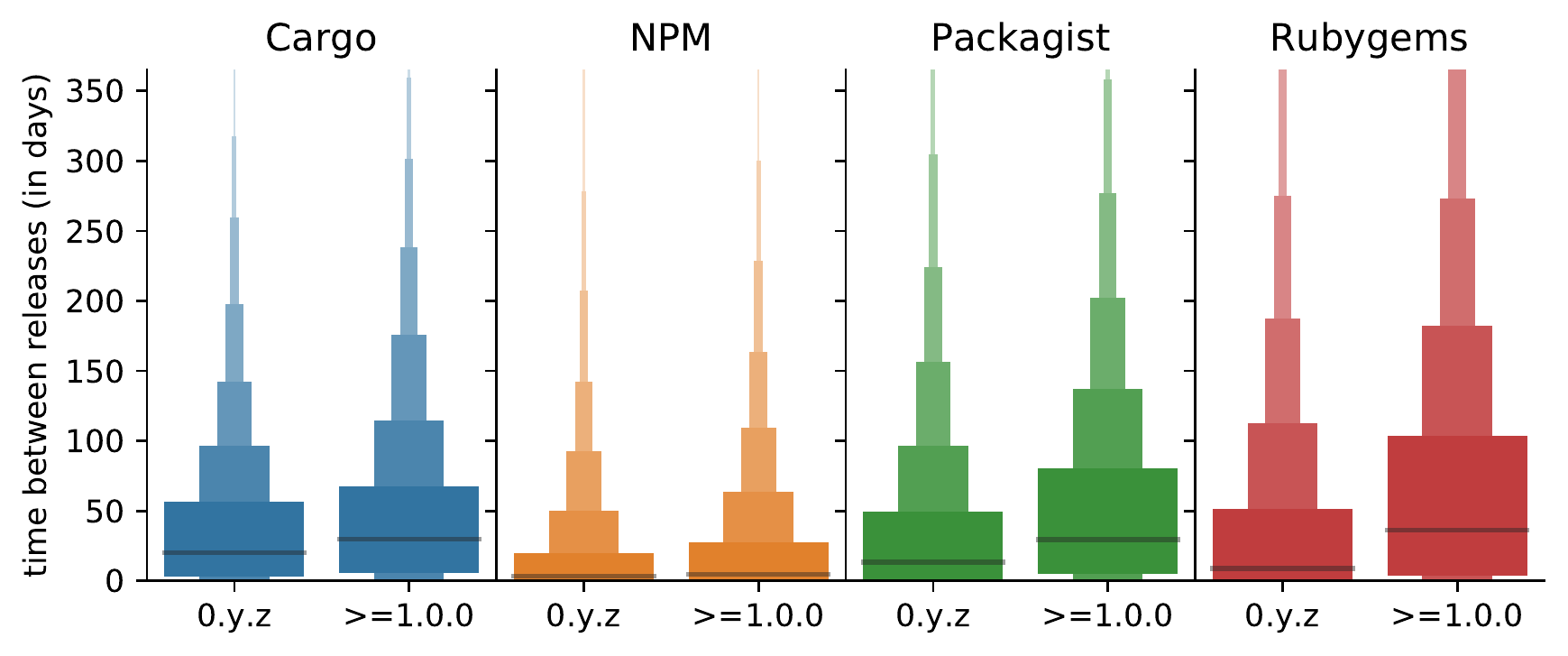}
    \caption{
        Distributions of the average \changed{number of days} between consecutive releases, for \preone and \postone releases.
    }
    \label{fig:rapid_update_delay}
\end{figure}

One would expect packages under initial development to \changed{publish new releases} more frequently than mature packages.
This is notably assumed by the \semver policy that states that ``\emph{major version zero is all about rapid development}''.
To verify this assumption we computed the distribution of the average time per package between consecutive releases \changed{(a.k.a. the period, corresponding to the reciprocal of the release frequency)}, for \preone and \postone releases respectively.
\fig{fig:rapid_update_delay} shows the boxen plots~\cite{letter-value-plot} for these distributions.

We observe that for all package distributions, \preone releases are more frequently \changed{published} than \postone releases (\ie the average time between releases is higher in \postone than in \preone releases).
For instance, for \cargo, the median was 20.2 \changed{days} for \preone and 29.9 \changed{days} for \postone releases. For \npm the median values were
3.4 and 4.9 days respectively, for \packagist 13.4 and 29.4 days, and for \rubygems 8.9 and 36.2 days.
To confirm the statistical significance of these differences between \changed{the release frequencies of} \preone and \postone releases, we carried out \changed{one-sided} Mann-Whitney-U tests~\cite{mann1947}.
The null hypothesis, stating that there is no difference between \changed{the release frequencies of} \preone and \postone releases, was rejected for all four package distributions with $p<0.01$ (adjusted after Bonferroni-Holm method to control family-wise error rate~\cite{10.2307/4615733}).
However, the effect size (measured using Cliff's delta $d$~\cite{Cliff1993494}) revealed that the observed differences were \emph{negligible} for \cargo ($|d| = 0.099$) and \npm ($|d| = 0.041$), and \emph{small} for \packagist ($|d| = 0.179$) and \rubygems ($|d| = 0.240$), following the interpretation of $|d|$ by Romano \etal~\cite{romano2006exploring}.

\changed{We also observe that the time to release a new version is lower in \npm than in the other package distributions, \ie \npm packages seem to publish releases more frequently.}
%
Mann-Whitney-U tests confirmed this observation with statistically significant differences ($p<0.01$), implying that \changed{releases} in \npm are indeed \changed{published} more frequently than \changed{releases} in \cargo, \packagist and \rubygems.
\changed{However, the effect size turned out to be \emph{small} in all cases:} $|d| = 0.327$ for \cargo, $|d| = 0.325$ for \packagist and $|d| = 0.185$ for \rubygems.
\changed{We also compared the release frequency of packages in \cargo, \packagist and \rubygems. While we found statistically significant differences between them, the effect size was always \emph{negligible} ($0.028 \le |d| \le 0.108$).}

\medskip

\begin{figure}[!htbp]
    \centering
    \includegraphics[width=\figsize]{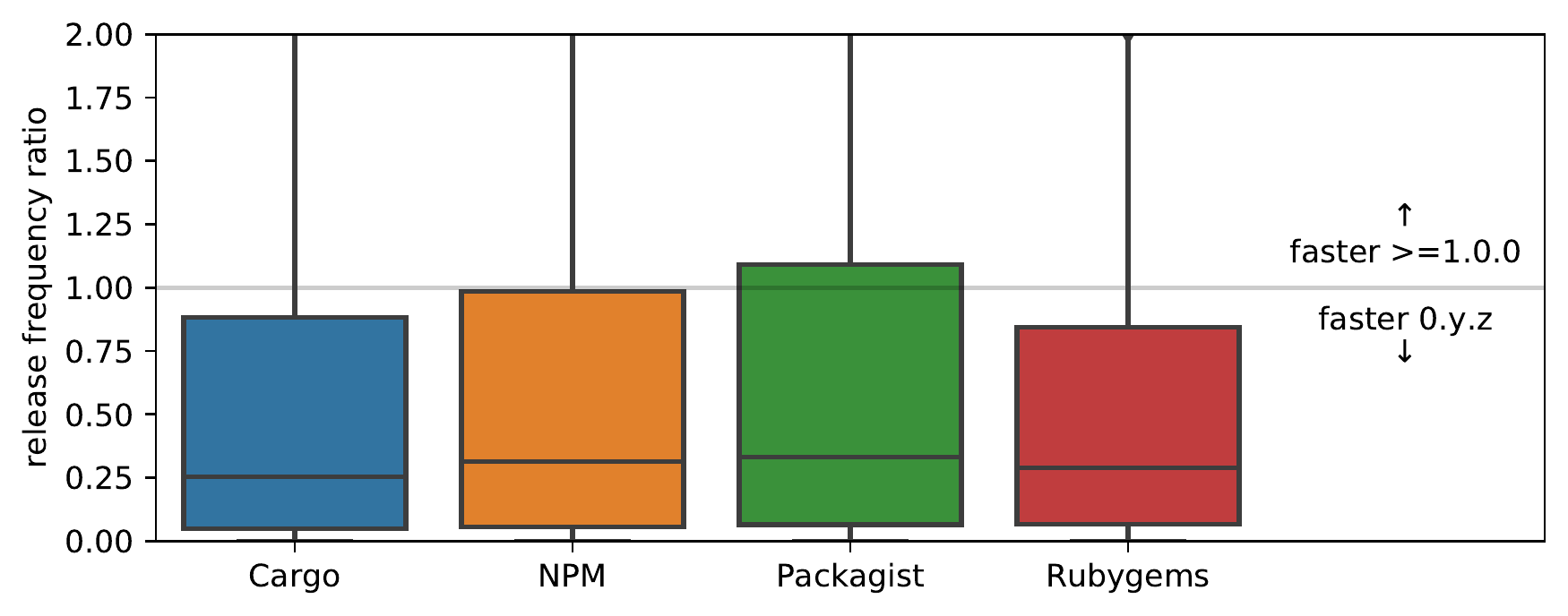}
    \caption{
        \changed{Distributions of the ratio between the release frequencies of  \postone and \preone releases for each package.}
    }
    \label{fig:rapid_update_ratio}
\end{figure}

\changed{
The previous analysis compared the release frequency between all \preone releases and all \postone releases of a {\em package distribution}, providing insights for the whole package distribution at once.
The following analysis focuses on the evolution of the release frequency between the \preone and \postone releases of {\em individual packages}.
For each package, we compared the frequency of its \preone releases to the frequency of its \postone releases.\footnote{The release frequency of \preone (resp. \postone) releases is obtained by dividing the number of \preone (resp. \postone) releases by the time between the first and last \preone (resp. \postone) releases.}

The boxplots in \fig{fig:rapid_update_ratio} show the distribution of the ratio between the release frequency of \postone and \preone releases of each package.
A ratio above 1 implies that the \postone releases of a package are more frequently published than its \preone releases, while a ratio below 1 means that the \preone releases of the package are more frequently published.
For obvious reasons, only packages having both \preone and \postone releases were considered for this analysis.
}

\changed{We observe that in all package distributions, the large majority of packages have a ratio below 1, indicating that \preone releases are published more frequently than \postone releases. The median ratio oscillates between 0.25 (for \cargo) and 0.33 (for \packagist), indicating that \preone versions are released 3 to 4 times more frequently than \postone releases.
There are only from 22\% (for \rubygems) to 26.7\% (for \packagist) packages whose \preone release frequency is lower or equal to their \postone release frequency.}

\begin{custombox}
    \preone releases are \changed{published} more frequently than \postone releases, but the effect is small for \packagist and \rubygems, and negligible for \cargo and \npm.
    \changed{The release frequency of most packages is higher for their \preone releases than for their \postone releases.}
    Regardless of their version number, package releases are \changed{published} more frequently in \npm than in the other distributions.
\end{custombox}
\subsection{Are \preone Package Releases Required by Other Packages?}
\label{sec:usage}

If one assumes that \preone packages are still under initial development, it could be considered unsafe to rely on them since they are expected to be less complete and less stable than production-ready packages.
This is confirmed by the \semver policy~\cite{semver2}: ``\emph{If your software is being used in production, it should probably already be 1.0.0.}''
Moreover, such \preone packages are likely to require extra effort from maintainers of packages depending on them.
Indeed, since ``\emph{anything may change at any time [and] the public API should not be considered stable}'', dependent packages are more likely to face breaking changes with \preone packages than with \postone packages.
\semver even recommends that ``\emph{If you have a stable API on which users have come to depend, you should be 1.0.0}''.

This research question therefore studies the extent to which packages rely on such \preone packages. 
\changed{Since the dependencies of a package can evolve over time, we consider the dependencies expressed in the latest available release of each package, hence reflecting the state of the latest snapshot in the package distributions.}
\tab{tab:usage} reports the proportion of dependent packages (\%~sources) relying on at least one \preone package, and the proportion of required packages (\%~targets) being used by at least one \postone package.
We distinguish between \preone and \postone sources and targets.

\begin{table}[!h]
    \caption{Proportion of source and target packages, depending on or required by \preone and \postone packages.}
    \label{tab:usage}
    \centering
    \begin{tabular}{cc|rr|rr}
        \bf \changed{package} & & \multicolumn{4}{c}{\bf target} \\
        \bf \changed{distribution} & \bf source & \bf \preone & \bf \postone & \bf \preone & \bf \postone\\
        \hline\hline
        \multirow{2}{*}{\cargo} & \bf \preone  & \cellcolor{gray!20} 82.0 & 9.9 & \cellcolor{gray!20} 76.9 & 5.9 \\
                                    & \bf \postone & \cellcolor{gray!20} 6.8 & 1.4 & \cellcolor{gray!20} 11.5 & 5.8 \\
        \hline
        \multirow{2}{*}{\npm} & \bf \preone  & \cellcolor{gray!20} 23.7 & 21.3 & \cellcolor{gray!20} 23.9 & 10.4 \\
                                  & \bf \postone & \cellcolor{gray!20} 19.8 & 35.2 & \cellcolor{gray!20} 14.6 & 51.1 \\
        \hline
        \multirow{2}{*}{\packagist} & \bf \preone  & \cellcolor{gray!20} 7.9 & 20.3 & \cellcolor{gray!20} 13.1 & 10.4 \\
                                        & \bf \postone & \cellcolor{gray!20} 5.5 & 66.3 & \cellcolor{gray!20} 6.1 & 70.4 \\
        \hline
        \multirow{2}{*}{\rubygems} & \bf \preone  & \cellcolor{gray!20} 17.4 & 56.7 & \cellcolor{gray!20} 20.9 & 34.7 \\
                                       & \bf \postone & \cellcolor{gray!20} 5.0 & 20.9 & \cellcolor{gray!20} 8.6 & 35.8 \\
        \hline\hline
        \multicolumn{2}{r}{proportionally to} & \multicolumn{2}{c|}{\bf \% sources} & \multicolumn{2}{c}{\bf \% targets}\\
    \end{tabular}
\end{table}

The reported proportions vary greatly from one package distribution to another.
For instance, a large majority ($88.8\%=82 + 6.8$) of the dependent packages in \cargo rely on at least one \preone release.
82\% of these dependent packages are \preone while only 6.8\% are \postone.
For \packagist and \rubygems, the inverse is true: a large majority of dependent packages ($86.6\%=20.3+66.3$ for \packagist, and $77.6\%=56.7+20.9$ for \rubygems) rely exclusively on \postone package releases.
\npm falls somewhere in the middle of both extremes, with 43.5\% ($=23.7+19.8)$ dependent packages relying on at least one \preone package release, and the remaining 56.5\% ($=21.3+35.2$) relying exclusively on \postone package releases.
Nevertheless, in all four package distributions there is still a large number of dependent packages relying on at least one \preone package release.

When considering these numbers proportionally to the set of required packages (\ie \% targets), we observe that 88.4\% ($=76.9+11.5$) of the required packages in \cargo are \preone packages.
This is in stark contrast with \packagist where this proportion is only 19.2\% ($=13.1+6.1$).
\npm and \rubygems fall somewhere in between, with 38.5\% ($=23.9+14.6$) of the required \npm packages being \preone packages,
and 29.5\% ($=20.9+8.6$) for \rubygems.
This indicates that in all four package distributions, at varying degrees, many \preone packages are still being used by other packages, including \postone ones.
This is rather counter-intuitive: even though common wisdom says that \preone packages are more likely to be unstable because they are still under inital development, package maintainers frequently depend on them.

\medskip

\begin{figure}[!htbp]
    \centering
    \includegraphics[width=\figsize]{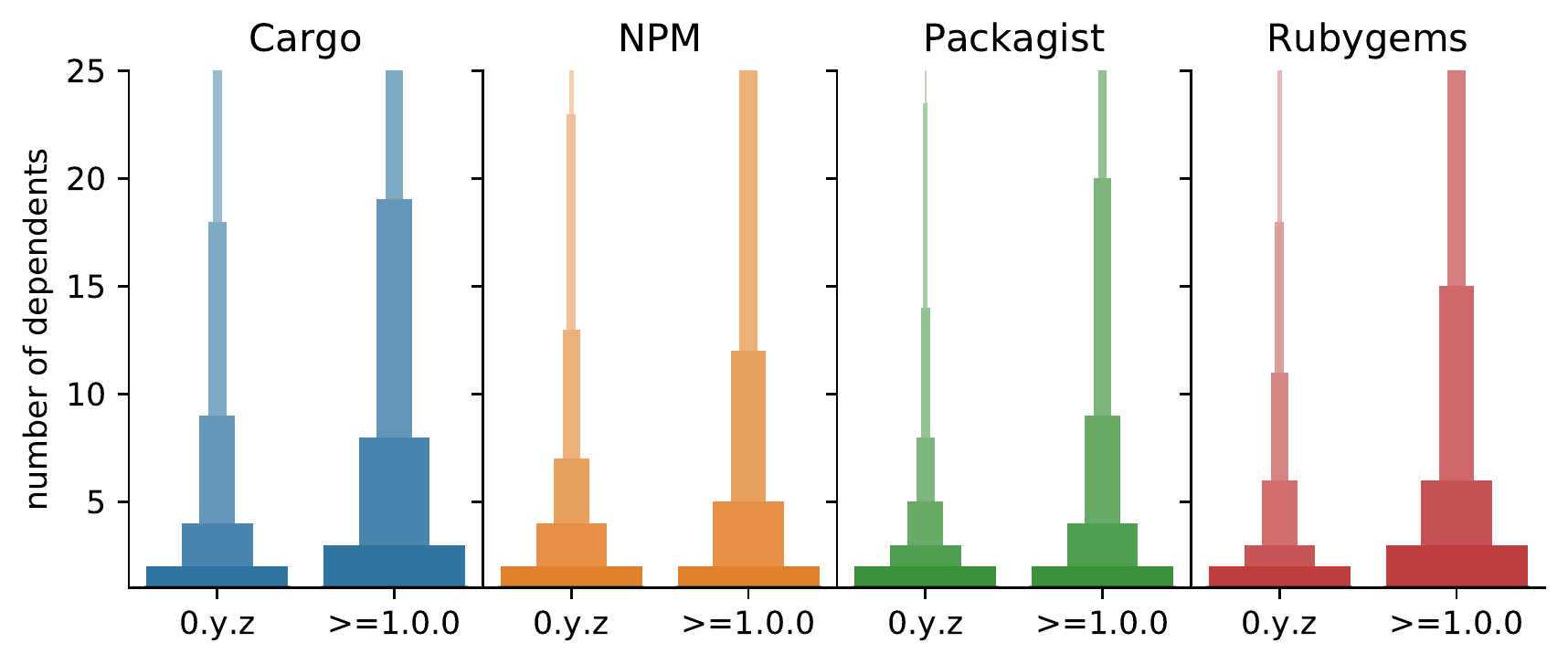}
    \caption{
        \changed{Distributions of the number of dependent packages for required \preone and \postone packages.}
        }
    \label{fig:usage_dependents}
\end{figure}

To \changed{determine} to which extent \preone and \postone packages are required, we computed for each package distribution their number of dependent packages (\ie reverse dependencies).
\changed{If a package has both \preone and \postone releases being required, we distinguish between packages depending on its \preone releases from packages depending on its \postone releases. We do so by looking at whether the highest release accepted by the dependency constraint is a \preone or a \postone release, complying with the default behaviour of the package managers used in the considered package distributions.}
\fig{fig:usage_dependents} shows the boxen plots of the distribution of the number of dependent packages for \changed{required packages, distinguising between packages depending on their \preone or on their \postone releases, respectively.}

Regardless of the considered package distribution, we only observe small differences between the number of dependent packages \changed{for} required \preone and \postone packages.
On average, the number of dependent packages for \preone packages oscillates between 2.1 (for \packagist) and 4.6 (for \cargo), while for \postone packages it is comprised between 6.9 (for \packagist) and 10.7 (for \cargo).
In all cases, the median number of dependent packages is $1$.
We statistically compared for each package distribution the number of dependent packages for \preone and \postone packages using a Mann-Whitney-U test to find evidence of a statistical difference.
The null hypothesis was rejected for all four package distributions ($p<0.01$ after Bonferroni-Holm correction), indicating that \postone packages are required more often.
However, the effect size was \emph{negligible} for all four package distributions ($0.066 \leq |d| \leq 0.108$).

\begin{custombox}
    Many packages are depending on \preone packages, ranging from 13.4\% of all dependent packages in \packagist to 88.8\% in \cargo.
    Many \preone packages are required by other packages, ranging from 19.2\% of all required packages in \packagist to 88.4\% in \cargo.
    There is no practical difference between the number of packages depending on \preone and \postone packages.\\
    \changed{Maintainers of \preone packages should strive to make their packages} cross the 1.0.0 barrier if they are  used by other production-ready packages.
\end{custombox}

\subsection{How Permissive Are Dependency Constraints Towards Required \preone Packages?}
\label{sec:constraints}

This research question focuses on a variation on the theme of unstable initial development packages.
Under the premise that \preone packages are unstable, the \semver policy assumes that any update of such a package could introduce backward incompatible changes: ``\emph{Major version zero (0.y.z) is for initial development. Anything may change at any time. The public API should not be considered stable.}''~\cite{semver2}

\changed{When specifying package dependencies, package maintainers make use of dependency constraints to specify which releases of the required package are allowed to be installed.
To ease the definition of such constraints in combination with \semver, package distributions provide specific notations to accept patches (\eg ${\sim}1.2.3$) or minor releases (\eg \caret$1.2.3$).
However, these notations are not the only way to specify dependency constraints (\eg ${>=}1.2.3, {<}2.0.0$ is equivalent to \caret$1.2.3$), and their interpretation is not always consistent across package distributions.\footnote{The reader is invited to consult Table~2 of \cite{decan2019tse} for more details.}
To take these differences into account, we wrote a dependency constraint parser\footnote{See \url{https://doi.org/10.5281/zenodo.4013419}} to convert the constraints using specific notations provided by each package distribution into a generic version range notation based on intervals~\cite{portion}.
For example, ${\sim}1.2$ is converted to the right-open interval $[1.2.0, 1.3.0)$ for \cargo and \npm, and to the right-open interval $[1.2.0, 2.0.0)$ in \packagist.
Based on this version range notation, we can easily identify whether a dependency constraint allows new patches, new minor and/or new major releases of the required package to be automatically installed.
For example, range $[1.2.0,1.3.0)$ accepts new patches because it contains version $1.2.x$ for $x > 0$, and does not accept new minor releases because it does not contain version $1.x.0$ for $x > 2$.
} 

\begin{figure}[!htpb]
    \centering
    \includegraphics[width=\figsize]{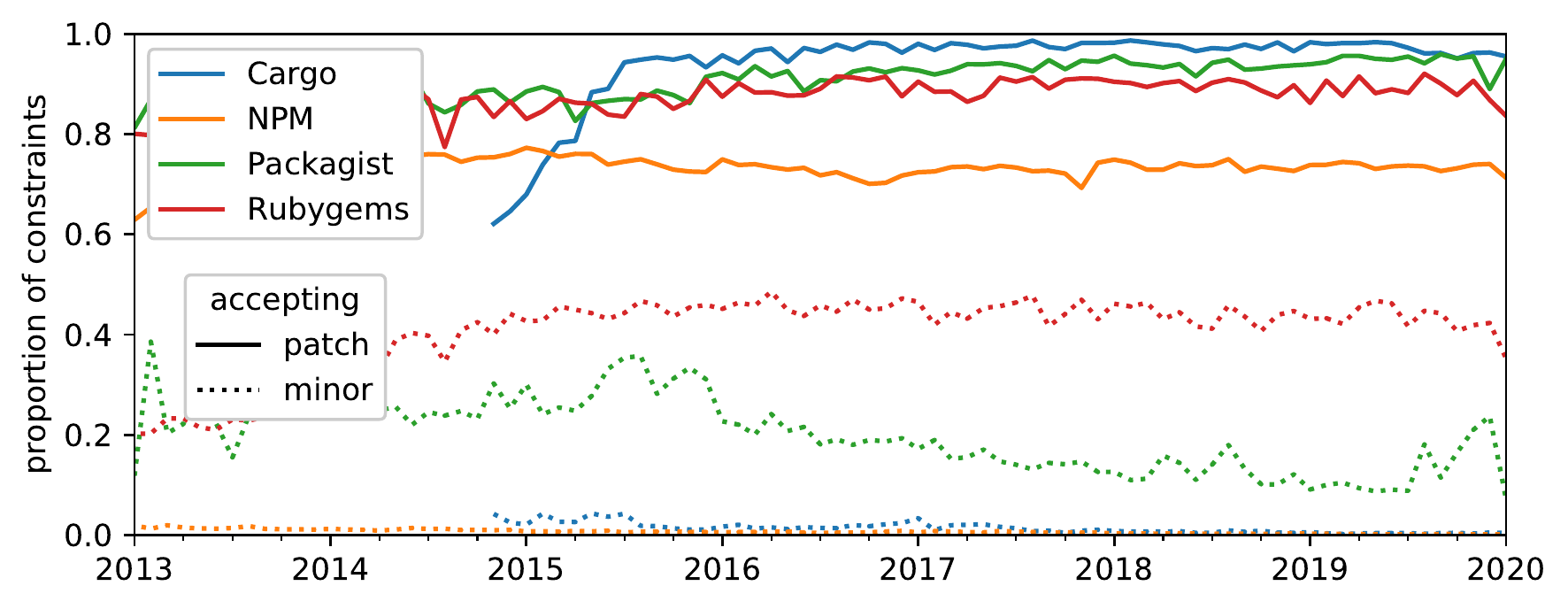}
    \caption{
        Evolution of the monthly proportion of dependency constraints accepting at least patches or minor \preone releases.
    }
    \label{fig:stable_permissiveness}
\end{figure}

For each considered package distribution we computed the monthly proportion of dependency constraints targeting \preone releases that accept at least patches or minor releases.
These proportions are shown in \fig{fig:stable_permissiveness} relative to the total number of constraints targeting \preone releases defined in newly distributed releases for each month.
\changed{To avoid the analysis to be biased by packages having many releases during the month, only the latest available release of each package was considered for each month.}

We observe that in all four distributions a large proportion of dependency constraints is accepting at least patches, from \changed{around 73\% for \npm to 94.2\% for \cargo}.
This proportion remains mostly stable over time. 
The proportion of dependency constraints accepting minor releases as well depends on the considered package distribution: it is close to zero for \cargo and \npm while, on the other hand, it fluctuates around \changed{40\%} in \rubygems.
\packagist sits in the middle of these extremes, with a steadily decreasing
proportion of constraints accepting minor releases since mid-2015, from \changed{33.1\%} to \changed{around 10\% in 2019}.

\medskip

\begin{figure}[!htpb]
    \centering
    \includegraphics[width=\figsize]{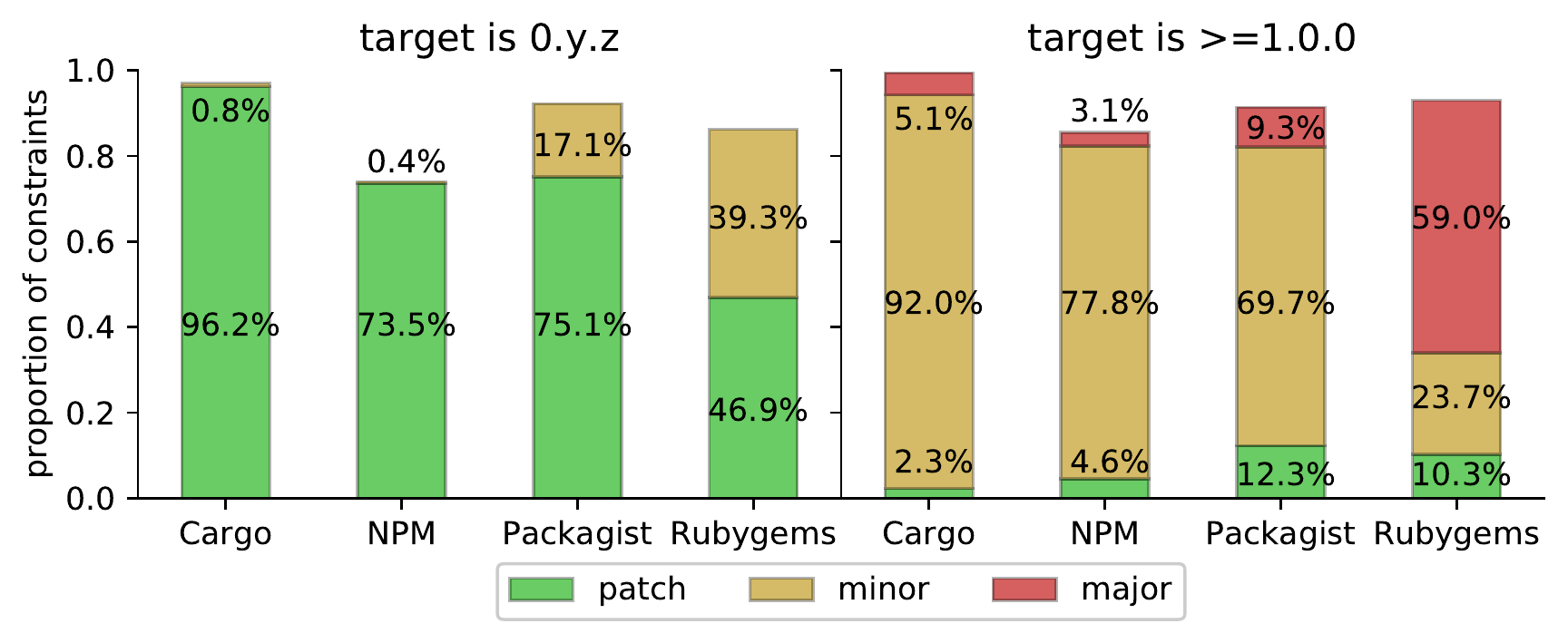}
    \caption{
        Proportion of dependency constraints accepting at most patches, minor or major releases, grouped by target releases.
    }
    \label{fig:stable_snapshot}
\end{figure}

Focusing on the latest snapshot of each package distribution, \fig{fig:stable_snapshot} reports the proportion of dependency constraints accepting at most patches, minor or major releases, separating between dependencies targeting \preone releases and \postone releases, respectively.

In order to be \semver-compliant, only strict constraints (\ie constraints that only accept a single version) to \preone package releases should be allowed to avoid the risk of introducing breaking changes.
However, we observe that the large majority of the dependencies targeting \preone releases are more permissive: they allow patches to be automatically installed, from 73.9\% (=$73.5+0.4$) in \npm to 97\% (=$96.2+0.8$) in \cargo.
\packagist and \rubygems are even more permissive, since a non-negligible proportion of dependency constraints to \preone package releases accept minor release updates as well (17.1\% of all constraints targeting \preone releases in \packagist, and 39.3\% in \rubygems).

For comparison, we carried out the same analysis for dependencies targeting \postone releases.
For those cases, the \semver policy considers it safe to accept minor release updates (since minor releases are expected to contain only backward compatible changes).
We indeed observe that the large majority of dependencies towards \postone releases accept minor releases as well.
For instance, we found 92.5\% ($=92 + 0.5$) of such dependencies in \cargo, 78.1\% ($=77.8+0.3$) in \npm, and 78.9\% ($=69.7+9.2)$ in \packagist.
For \rubygems, we found 82.7\% ($=23.7+59$) of such dependencies, a consequence of the presence of 59\% of dependencies accepting major releases as well.
For \rubygems, the high proportion of dependencies accepting minor \preone releases and major \postone releases indicates that its packages are not \semver-compliant, not even for \postone releases~\cite{decan2019tse}.

\begin{custombox}
    Most dependencies towards \preone releases accept new patches, indicating that these patches are expected to be backwards compatible. As such, the considered package distributions adopt a policy that is more permissive than \semver for \preone releases.
    \packagist and \rubygems are even more permissive, since more than one dependency constraint out of six also accepts minor releases.\\
    This relaxation w.r.t. \semver should be made explicit, or the \semver policy should be loosened to allow package maintainers to specify backwards compatible updates for \preone releases.
\end{custombox}
\subsection{Do \github Repositories for \preone Packages Have Different Characteristics?}
\label{sec:characteristics}

So far, we have not really been able to discern any difference between \preone and \postone packages.
\changed{If we assume that \postone packages are more production-ready, stable and mature than \preone packages, we can expect their git repositories to have more stars, more forks, more contributors or less open issues than the git repositories of \preone packages.}
In this research question, we aim to compare \preone and \postone packages based on the characteristics of their git repository, focusing on the ones being hosted on \github, the most popular distributed collaborative development platform.

Since not all packages have an associated repository on \github, \tab{tab:repositories} reports only the proportion of \preone and \postone packages that have \changed{a known \github repository in our dataset.}

\begin{table}[!h]
    \caption{Proportion of packages with a known \github repository.}
    \label{tab:repositories}
    \centering
    \begin{tabular}{c|cc|c}
        \bf distribution & \bf \preone packages & \bf \postone packages & \bf all packages\\
        \hline\hline
        \cargo & 72.1\% & 85.9\% & 73.1\%\\
        \npm & 62.5\% & 59.8\% & 61.0\%\\
        \packagist & 94.8\% & 94.3\% & 94.4\%\\
        \rubygems & 65.3\% & 70.9\% & 66.7\%
    \end{tabular}
\end{table}

We observe that a large majority of all packages have an associated repository on \github, from 61\% for \npm to 94.4\% for \packagist.
This higher proportion for \packagist is a consequence of the way packages are made available. Indeed, to submit a new package on \packagist, one has to provide the URL of a public repository.
We also observe a slight difference between the proportion of \preone and \postone packages having a repository on \github: there are proportionally more repositories for \postone packages in \cargo (85.9\% versus 72.1\%) and in \rubygems (70.9\% versus 65.3\%) while there are proportionally more repositories for \preone packages in \npm (62.5\% versus 59.8\%) and in \packagist (94.8\% versus 94.3\%).

\medskip

\begin{figure}[!htbp]
    \centering
    \includegraphics[width=\figsize]{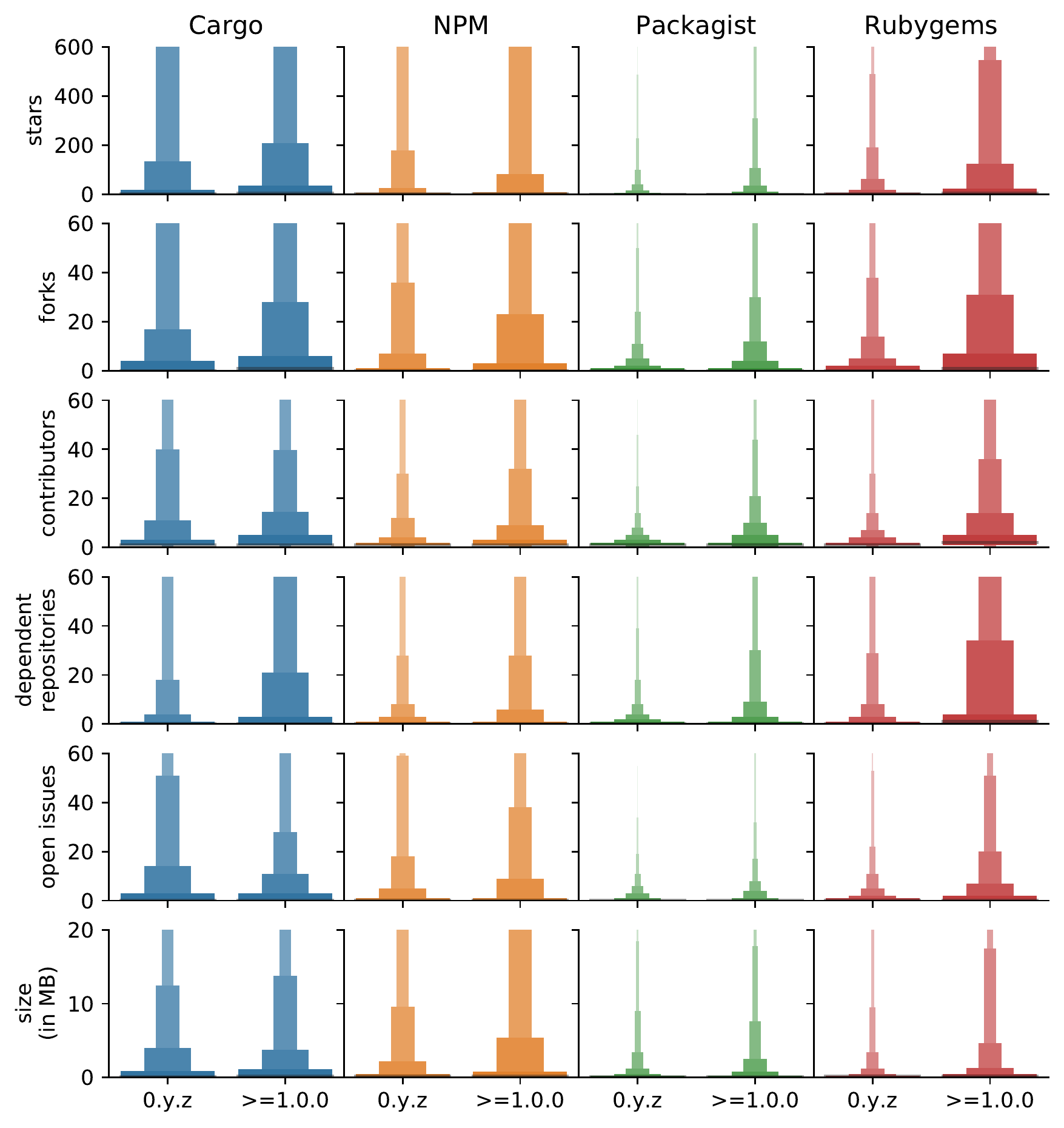}
    \caption{
        Distributions of the number of stars, forks, contributors, dependent repositories, \changed{open issues} and git repository size (in MB) for \preone and \postone packages.
    }
    \label{fig:char_distribution}
\end{figure}

\changed{For the \github repository associated to each package, we used our dataset to extract the number of stars, forks, contributors, open issues and the size of the git repository (expressed in megabytes). We also extracted the number of dependent repositories, \ie software projects developed in \github repositories and that depend on the given package.}
\fig{fig:char_distribution} shows the boxen plots of the distributions of these characteristics for each package distribution, distinguishing between repositories hosting \preone and \postone packages.

We observe a slight difference between \preone and \postone repositories for all \changed{characteristics and for all} package distributions.
\changed{The differences are especially visible in \rubygems.}
We carried out Mann-Whitney-U tests between \preone and \postone package repositories for each characteristic to confirm these differences.
The null hypothesis stating that there is no difference between \preone and \postone package repositories was consistently rejected in all cases ($p<0.01$ after Bonferroni-Holm correction), \changed{with the notable exception of open issues in \cargo.}
\changed{This confirms} that \postone package repositories have more stars, forks, contributors, dependent projects, and open issues \changed{(except in \cargo)}. Moreover, they are larger than \preone package repositories.
Nevertheless, the effect size turned out to be \emph{negligible} for all comparisons in \cargo, \npm and \packagist ($0.006 \leq |d| \leq 0.132$) and \emph{small} in \rubygems ($0.156 \leq |d| \leq 0.258$).

\begin{custombox}
    A large majority of \preone and \postone packages have an associated repository on \github.
    Repositories for \postone packages have slightly more stars, forks, contributors, dependent projects \changed{and open issues (except in \cargo)}, but the \changed{differences are} negligible for \cargo, \npm and \packagist, and small for \rubygems.
\end{custombox}

\section{Discussion}
\label{sec:discussion}


\changed{This section aims to discuss about the lessons learned from the empirical analysis, as well as the recommendations that could be made to package maintainers and managers of package distributions based on these findings.}

\subsection{\changed{Qualitative evidence}}
\label{sec:qual-poll}

\changed{The results reported in this paper were based on quantitative evidence relying on historical analysis of package management data. To complement this quantitative analysis, we conducted a single question poll on \textsf{Twitter} and \textsf{LinkedIn} 
in December 2020.
The poll aimed to obtain insights in whether practitioners effectively consider it risky to depend on packages that are still in version \preone. The single multiple-choice question was:
\begin{quotation}
    \emph{``As a software developer, you need to depend on an open source package distributed through some package manager (like npm, Maven, Cargo, Packagist, RubyGems). Would you trust depending on a package with major version 0 (\eg version 0.5.1)?''}
\end{quotation}

\begin{table}[htbp]
   \centering
   \caption{Number of answers for each of the 4 possible responses to the multiple-choice question.}   \begin{tabular}{l|c} 
      \bf response    & \bf \# answers \\
      \hline\hline
      \em No      & \cellcolor{gray!7} ~7 \\ 
      \em Only if there is no alternative    & \cellcolor{gray!21} 21 \\ 
      \em Only after checking   & \cellcolor{gray!45} 45 \\ 
      \em Sure       & \cellcolor{gray!37} 37 \\ 
      \hline
      \multicolumn{1}{r}{Total} & 102\\
   \end{tabular}
   \label{tab:poll}
\end{table}

We received 102 responses in total, of which 58 on LinkedIn and 44 on Twitter. The results are summarised in \tab{tab:poll}.
From these results we conclude that the large majority of respondents considers it risky to depend on \preone package releases. This corresponds to the common convention that \preone packages are assumed to be under initial development and therefore potentially less stable than \postone releases.}

\begin{custombox}{\bf Lessons learned.}
\changed{
Common belief suggests that \preone package releases are less stable than \postone ones.
64\% of respondents perceive depending on \preone package releases as risky: they would prefer not to depend on \preone packages, or only if there is no alternative or after doing additional checks on that package.
}
\end{custombox}

\subsection{The 1.0.0 barrier}

\changed{The results we obtained from the quantitative analysis in this paper seem to be in shrill contrast with the common wisdom that was confirmed by the qualitative analysis presented in \sect{sec:qual-poll}.
We observed little difference between \preone and \postone packages, implying that many \preone packages can be assumed to be production ready and safe to use.
Given this little difference, a kind of psychological barrier related to  version 1.0.0 could perhaps explain why so few packages reach a \postone release.} A major version 1 is usually associated with the promise of a stable API and a mature library. For \npm, a developer witnesses that \emph{``[he] tends to associate 1.0.0 with a finished project, including tests, documentation, a nice landing page, and a lot of sample code.''}. However, he agrees that \emph{``while all of that is certainly important in its own right, these extra components of a project actually have nothing to do with the version number''}~\cite{JeremyKahnBlog2013}.

We believe that most developers of packages in the 0 version space avoid to cross the 1.0.0 barrier in order to keep the freedom to make API (breaking) changes, and to not have to commit to the (overly optimistic and unrealistic) bug-free nature of \postone releases, even if their package already reached this degree of maturity.
Because of this, developers have to rely on other ways to assess the maturity of \preone packages, as witnessed by another \rust developer who \emph{``[has] to go read the code to see if this is a `0.1.0' package which is basically finished, or a `0.1.0' full of partially implemented functionality''}~\cite{RedditFear100}.

It is not unusual that the decision to cross the 1.0.0 barrier is based on other factors, such as providing more functionalities or polishing the package.
For example, Tom Augspurger from the \textsf{pandas} developer team testifies that \emph{``\textsf{pandas} has been `production ready' for a while now, in the sense that it’s used in production at many institutions. But we still had a few major items we wanted to iron out before calling 1.0''}.\footnote{\url{https://jaxenter.com/python-pandas-1-0-0-tom-augspurger-167593.html}}
Similarly, Piotr Solnica, maintainer of \textsf{dry-validation}, one of the most used packages in \rubygems, explains \emph{``[they] have got a long backlog in the issue tracker and [he] would like to address all of those issues ASAP and make the codebase simpler AND add features that are even more powerful than what we have already. The final goal is to turn this into 1.0.0 in a couple months from now.''}.\footnote{\url{https://discourse.dry-rb.org/t/plans-for-dry-validation-dry-schema/215}}

There are plenty examples of popular \preone packages being developed for years, known for their stability and maturity and being used in production by thousands of users.
One such example is \textsf{style-loader}, one of the most famous libraries on \npm.
Despite its widespread use (it has more than 10K dependent packages, and several millions weekly downloads), it only reached its first \postone release in August 2019, after more than 7 years of development and 54 releases.
Another example is the \textsf{syn} package in \cargo. Its first \postone release was reached after 3 years of development and 122 releases. At that time, it was already used by one third of all packages in \cargo. For its developers, the decision to release a 1.0.0 version \emph{``signifies that \textsf{Syn} is a polished and stable library and that it offers a user experience we can stand behind as the way we recommend for all Rustaceans to write procedural macros''}.\footnote{\url{https://github.com/dtolnay/syn/issues/687}}

It is not surprising that we did not observe any fundamental difference between \preone packages and \postone packages, both in terms of update frequency and usage by other packages, even if the common belief suggests such a difference holds.
This belief is reinforced by how package distributions and versioning policies treat both types of packages.
Indeed, we found that \npm, \packagist and \cargo make an explicit distinction between how dependency constraints are treated for \preone and \postone releases, based on the presumed degree of maturity.

An alternative approach consists of not associating a different versioning policy to \preone and \postone releases.
This is the case for \haskell packages, for which the official versioning policy 
explicitly states that ``\emph{packages with a zero major version provide the same contractual guarantees as versions released with a non-zero major version}''.\footnote{\url{https://pvp.haskell.org/faq/}}
This does not seem to be a perfect solution either, since in practice it does not encourage maintainers to cross the 1.0.0 barrier either: ``\emph{an easily spottable plague of an absolute majority of \haskell packages is that they get stuck in the 0.x.x version space, thus forever retaining that `beta' feeling even if the package's API remains stable for years and has dependencies counted by thousands}''.\footnote{\url{https://www.reddit.com/r/haskell/comments/31e3jj/}}

\begin{custombox}{\bf Lessons learned.}
    Most packages remain stuck in the zero version space.
    There \changed{seems to be} a psychological barrier associated to version 1.0.0. The use of different rules and versioning policies for \preone and \postone releases by the package distributions only \changed{seems to} reinforce this barrier.

    \medskip\noindent{\bf Recommendations.}
    There is no fundamental reason why \preone releases should not fulfil the same contracts and promises as \postone releases, especially as soon as a package is being distributed and used by others.
    Package maintainers should \changed{strive to cross the 1.0.0 barrier, especially if their packages} have been developed for years and are indubitably ready for production.
\end{custombox}

\medskip

\changed{
\subsection{Maturity of distributed packages}

Our analysis found little difference between \preone and \postone packages, implying that many \preone packages can be assumed to be production-ready and safe to use.
This could be explained by the fact that, as mentioned by a \rust developer, \emph{``publishing already implies some degree of production readiness''}~\cite{RedditFear100}.
Although it is probably true that packages being available in package distributions are at least functional and offer a minimum of features, none of the four consider package distributions actually require their distributed packages to achieve some degree of production readiness.

We have not found any guidelines in the documentation of these package distributions specifying or suggesting the minimal requirements a package must or should fulfil for its distribution.
At most, the documentation of \rubygems mentions that \emph{``testing your gem is extremely important''} and that \emph{``developers tend to view a solid test suite (or lack thereof) as one of the main reasons for trusting that piece of code.''}\footnote{\url{https://guides.rubygems.org/make-your-own-gem/}}

We are convinced that package distributions and package developers would greatly benefit from guidelines defining the minimum requirements for a package to be distributed.
For example, \textsf{CRAN}, the official package distribution for the \textsf{R} programming language, states that \emph{``CRAN hosts packages of publication quality and is not a development platform. A package’s contribution has to be non-trivial.''}
It imposes several requirements with the respect to the quality and maturity a package must have in order to be accepted for distribution.\footnote{See \url{https://cran.r-project.org/web/packages/policies.html} for more details.}

The lack of guidelines in the considered package distributions means that anyone can basically publish anything in these package distributions, including immature, unstable, incomplete or even non-working packages.
This explains why we found packages with clearly deviating and undesirable behaviour in our dataset, as explained in \sect{sec:data}.

Imposing requirements and/or responsibilities on the packages that can be distributed in a package distribution could prevent many packages from being ``officially distributed''.
For example, a maintainer may not be willing or able to meet these requirements, or may not be willing to commit to certain responsibilities, especially if they are not limited in time.
However, just because a package cannot or is not distributed on a package distribution does not prevent other packages or users from making use of it.
Indeed, the package manager tools of the considered package distributions all provide an easy way to install or depend on packages from another source (\eg from a private package distribution, from a git repository, etc.).
}

\begin{custombox}{\bf Lessons learned.}
    \changed{None of the considered package distributions provide guidelines on the minimal requirements that a package should fulfil to be distributed.}

    \medskip\noindent{\bf Recommendations.}
    \changed{Package distributions should clarify the expectations and responsibilities associated with distributing a package through the package manager.}
\end{custombox}

\medskip

\changed{\subsection{Version policies of package distributions}}

We found in \sect{sec:prevalence} that \npm is the only one of the considered distributions that exhibits a clearly decreasing proportion of \preone packages, starting from April 2014.
At that time, \npm \changed{changed} the initial version of packages created through \texttt{npm init} to 1.0.0 instead of 0.1.0.\footnote{See \url{https://github.com/npm/init-package-json/commit/363a17bc3}}

Based on this observation, we conjecture that policies and tools impact the proportions of \preone packages in package distributions.
For instance, the \texttt{cargo init} command to create new packages for \cargo sets the initial version of a new package to 0.1.0.
Similarly, while \textsf{gem}, the official package manager for \rubygems, does not allow to create new packages, \textsf{bundler}, its recommended alternative,\footnote{See \url{https://guides.rubygems.org/make-your-own-gem/}} sets the initial version of a newly created package to 0.1.0.
On the other hand, \textsf{composer}, the official package manager for \packagist, does not specify a default initial version for newly created packages for \packagist since version numbers in \packagist are deduced from git tags.\footnote{See \url{https://getcomposer.org/doc/articles/versions.md}}

To understand to which extent the default initial version affects the considered package distributions, we looked at the version numbers adopted by the first public release of new packages.
\tab{tab:prevalence_initial} reports about these version numbers and the proportion of packages created in 2019 that adopted them for their first public release.

\begin{table}[!h]
    \caption{Proportion of packages created in 2019 in function of the version number adopted for their first public release.}
    \label{tab:prevalence_initial}
    \centering

    \begin{tabular}{c|rrrr|rr}
        \multicolumn{5}{c}{} & \multicolumn{2}{c}{\bf other}\\
        \bf distribution & \bf 1.0.0 & \bf 0.1.0 & \bf 0.0.1 & \bf 0.0.0 & \bf \preone & \bf \postone\\
        \hline\hline
        \cargo & 2.9\% & \cellcolor{gray!20} 55.3\% & 7.5\% & 13.5\% & 19.6\% & 1.3\%\\
        \rubygems & 11.6\% & \cellcolor{gray!20} 47.1\% & 13.6\% & 4.7\% & 14.3\% & 8.8\%\\
        \npm  & \cellcolor{gray!20} 40.8\% & 11.5\% & 18.0\% & 2.9\% & 10.7\% & 16.1\%\\
        \packagist & \cellcolor{gray!20} 48.5\% & 13.3\% & 12.2\% & 0.6\% & 7.5\% & 18.0\%\\
    \end{tabular}
\end{table}

We observe that \cargo and \rubygems, the two distributions that set the default initial version of a package to 0.1.0, have a much higher proportion of packages being released with a 0.1.0 initial version number.
On the other hand, \npm and \packagist have a much higher proportion of packages being released with a 1.0.0 initial version number.
Specifically for \npm, we computed the proportions for packages having been first distributed in 2013, the year preceding the adoption of 1.0.0 as the default initial version.
We found that only 6.9\% of packages were first released with 1.0.0, while 24\% were released with 0.1.0, and 37.2\% with 0.0.1.
This seems to confirm that the adoption by \npm of a new default value for the initial version of a package impacted the version number used by newly created packages.

\changed{However, while the adoption of this policy seems to have pushed maintainers to select a \postone version number for the initial version of their packages, nothing indicates that these packages are more mature or more stable than \preone packages.
Even if nothing in this policy prevents a package maintainer to choose a \preone version number for the initial version of a package, a maintainer may also opt for a \postone version regardless of the maturity and stability of the package.
Pushing developers to adopt a \postone version without clarifying the expectations and responsibilities associated with such versions is not an acceptable solution. It may prevent developers from communicating to dependent packages that their packages are, in fact, immature and unstable.}


\begin{custombox}{\bf Lessons learned.}
\changed{Different package distributions adopt different versioning policies,
explaining the observed differences for the proportion of \preone packages.}

    \medskip\noindent{\bf Recommendations.}
    \changed{\cargo and \rubygems should adapt their versioning policies} to incite package maintainers to move out of the zero version space, the same way as \npm has successfully done in 2014.
    \changed{However, package distributions should first clarify the expectations and responsabilities associated with a 1.0.0 version.}
\end{custombox}

\medskip

\subsection{Semantic Versioning}

Our findings revealed that the \semver policy does not correspond to how \cargo, \npm and \packagist deal with \preone package releases in practice. This difference can be quite confusing for practitioners.
Firstly, while \semver considers that ``\emph{major version zero is all about rapid development}'', we found no conclusive evidence of this.
Indeed, only a small proportion of packages crossed the 1.0.0 barrier, even after years of development.
Moreover, we observed that \preone releases are not updated considerably more frequently than \postone releases.

Secondly, \semver has no specific rule dictating how to increment the version number of a \preone release to indicate a compatible update. The policy is overly restrictive by assuming that ``\emph{anything may change at any time}'' and that ``\emph{the public API should not be considered stable''}.
\changed{Because of this, developers of packages adhering to \semver cannot convey the backwards compatibility of \preone releases through their version numbers. As a consequence, there is no way for a developer of a dependent package to declare a dependency constraint such that backwards compatible releases are allowed while at the same time preventing incompatible ones to be adopted. This means they have to decide either to face the risk of breaking changes, or to stay on the safe side by preventing the automatic installation of new \preone releases and not being able to benefit from the bug and security fixes of these releases.}
Package distributions have therefore introduced notations and guidelines to circumvent this restriction.
For example, \cargo defines caret requirements (\ie \caret x.y.z) as a way to ``\emph{allow \semver compatible updates to a specified version}'' but its implementation accepts patches for \preone releases.\footnote{\url{https://doc.rust-lang.org/cargo/reference/specifying-dependencies.html}}
The same notation also exists in \npm and \packagist and, although their semantics complies with \semver for \postone releases, it does not for \preone releases as it allows patches.
Another example stems from the documentation of \npm that recommends ``\emph{starting your package version at 1.0.0 to help developers who rely on your code}''\footnote{\url{https://docs.npmjs.com/about-semantic-versioning}} and even explicitly mentions that ``\emph{many authors treat a 0.x version as if the x were the ``breaking-change'' indicator}''.\footnote{\url{https://docs.npmjs.com/misc/semver}}
\rubygems is even more permissive since, while its official documentation ``\emph{urges gem developers to follow the Semantic Versioning standard}''\footnote{\url{https://guides.rubygems.org/patterns/\#semantic-versioning}}, it makes no difference between \preone and \postone when detailing how to increment version numbers with respect to backwards compatibility.
This is a likely explanation for the findings in \sect{sec:constraints} that most dependencies towards \preone releases accept new patches.

\begin{custombox}{\bf Lessons learned.}
    Package maintainers in the considered package distributions do not strictly follow \semver for \preone releases, and adopt a more permissive policy.

    \medskip\noindent{\bf Recommendations.}
    \changed{To fully benefit from \semver, maintainers of mature packages should use \postone version numbers.}
    \changed{Package distributions should explicitly document and communicate any deviation from the \semver policy}, or the \semver policy should be adapted to allow maintainers to specify backwards compatible updates for \preone releases.
\end{custombox}

\changed{\subsection{Wisdom of the crowds}}

The \semver policy also considers that ``\emph{if you have a stable API on which users have come to depend, you should be 1.0.0}'' and that ``\emph{if your software is being used in production, it should probably already be 1.0.0''}.
This is especially relevant in package distributions such as \cargo in which it is usual to have public dependencies as part of the API, as witnessed by a \rust developer: \emph{``You may be ready to release 1.0 but if your crate reexports items from another and that crate hasn't released its 1.0 yet, then you're talking about entering a new level of dependency hell where you can't even upgrade dependencies without publishing a new major release}''~\cite{RedditFear100}.

\changed{However, our quantitative evidence reveals this \semver guideline is not strictly followed in practice.}
Many \preone packages are heavily used by other packages, including by ``production-ready'' (\ie~\postone) packages.
For example, the \textsf{axios} package on \npm has not yet reached a \postone release, even though it 
is directly required by 30K other \npm packages, and it exceeds 5M weekly downloads.
A similar example for \cargo is the \textsf{rand} package.
It has more than 25M downloads and more than 3K direct dependent packages, despite still being in \preone since 2015 and having released more than 60 versions.

\changed{The fact that so many packages depend on \preone packages suggests that these \preone packages are more stable or more mature than what their version number suggests, hence reducing the risk of depending on them.
The wisdom of the crowds principle has already been proposed to assess the backwards compatibility of packages based on the permissiveness of their (reverse) dependency constraints~\cite{decan2019tse}.
In a similar vein, developers desiring to depend on a \preone package could rely on the number of existing dependents to assess the perceived maturity of  that package.
Package distribution managers could even go one step further, and provide automated support to suggest maintainers of such packages to upgrade their package version number to \postone based on this perceived maturity.
}

\begin{custombox}
{\bf Lessons learned.}
\changed{Many \preone packages are heavily used by other packages, even if \semver signals it is risky to do so. This suggests that those packages are more mature than what their version number reflects.}

\medskip\noindent{\bf Recommendations.}
\changed{Developers desiring to depend on \preone packages could rely on wisdom of the crowds to assess the risk of doing so. Package distributions could use the same principle to provide automated support for recommending package maintainers to cross the 1.0.0 barrier.
}
\end{custombox}

\section{Threats to Validity}
\label{sec:threats}


We discuss the main threats that may affect the validity of our findings, following the structure recommended by Wohlin \etal~\cite{Wohlin2000book}.

Threats to \emph{construct validity} concern the relation between the theory behind the experiment and the observed findings. 
%
The accuracy of our findings assumes that the package dependency metadata extracted from \textsf{libraries.io} is correct.
We checked this assumption in previous work that relied on the same dataset~\cite{DecanEMSE2018, decan2019tse}\changed{, by manually looking at hundreds of examples, as well as by comparing a subset of this dataset with the data available from the official package distributions (\eg the \npm package registry) or from \github.
Our findings may also be influenced by the ``noise''} present in the original data provided by the package distributions. 
As explained in \sect{sec:data}, we removed such noise by excluding package releases from \npm and \packagist that did not correspond to real development.
Another source of imprecision is caused by the preparatory step to convert dependency constraints to more generic version range notations, as explained in \sect{sec:constraints}.
This is unlikely to affect our results since the large majority of constraints (98.2\%) could be parsed.

Threats to \emph{internal validity} concern choices and factors internal to the study that could influence the observations we made.
We did not find any such threats in our work.

Threats to \emph{conclusion validity} concern the degree to which the conclusions we derived from our data analysis are reasonable.
Given that our findings are based on empirical observations and on statistical tests with a high confidence level ($\alpha = 0.01$ adjusted after Bonferroni-Holm method to control family-wise error rate~\cite{10.2307/4615733}), they are not affected by such threats.

The threats to \emph{external validity} concern whether the results can be generalized outside the scope of this study.
The proposed approach is certainly generalizable to other package distributions since it is mainly observational, as witnessed by the addition of \rubygems compared to our previous work~\cite{Decan2020Zero}.
The observed findings themselves, however, are specific to the considered package distributions, since they are highly dependent on their policies, practices and tools.
We already found important differences among the four package distributions we analysed, and we expect to see more such differences in other distributions, especially the ones relying on other versioning schemes (\eg \textsf{Hackage} for \textsf{Haskell} or \textsf{PyPI} for \textsf{Python}).
\section{Conclusion}
\label{sec:conclusion}

In order for a mature software project to be considered healthy, it should avoid depending on unstable and immature reusable packages that are still in their initial development phase.
A popular convention is to assume that \changed{\preone package releases} are more likely to be less complete, mature and stable than \postone package releases. \changed{This common belief is supported by anecdotal qualitative evidence that \preone releases are perceived more risky and should be used with care. This belief is reflected in} the versioning policies of package distributions that define different rules for \preone and \postone package releases.

\changed{This paper aimed at finding quantitative evidence of} how \preone and \postone package releases behave in the \cargo, \npm, \packagist and \rubygems package distributions.
We observed that \preone packages are prevalent in all four distributions, even contributing to more than 90\% of all packages in \cargo. \preone packages are as active as \postone packages, and they represent a large proportion of all package updates.
We found that only a small proportion of packages went from a \preone to a \postone release.
While the majority of them took a few months and a few updates to do so, more than one out of five packages needed more than a year to reach a \postone release.
We observed that \preone releases are \changed{published} slightly more frequently than \postone packages, but the difference is small in \packagist and \rubygems, and even negligible in \cargo and \npm. \changed{We also found that the release frequency of most packages is higher for their \preone releases than for their \postone releases.}

We found that many \preone packages are already used by other packages, and that many \postone packages are relying on \preone packages.
We studied how often \preone and \postone releases are required by other packages but found no practical difference between them.
We assessed whether \preone releases comply with the \semver policy by analysing dependency constraints towards \preone releases.
We found that the considered package distributions adopt a policy that is more permissive than \semver, since most of these dependencies accept new patches.
Finally, we found that a large majority of \preone and \postone packages have an associated repository on \github. Repositories related to \postone packages tend to have slightly more stars, more forks, more contributors, \changed{more open issues}, more dependent projects and are slightly larger than the ones related to \preone packages, especially for \rubygems.

\changed{These quantitative findings go against the common wisdom that \preone package releases should be used differently from \postone releases, as they are more likely to correspond to packages with a lower degree of maturity and stability usually associated to the initial development phase.
This observed discrepancy seems to come from the fact that many mature and production-ready packages remain stuck in the zero space.
To fully benefit from \semver they should be incited to upgrade to a \postone release. This can be done by package distribution managers, by adopting their versioning policies and by relying on wisdom of the crowds to detect which \preone releases are likely to be stable and ready to move out of the zero space.
Of course, the release number by itself is not sufficient to assess package maturity: there is a need for package distributions to clarify the expectations and responsibilities associated with distributing a package through the package manager.}

The presented research can be extended in many ways.
For example, one could rely on the development history of a package to assess at a fine level of granularity whether \preone releases actually correspond to rapid development (\eg based on the number and size of commits and code changes), contain less or less stable features (\eg based on the number of feature and pull requests), or are more prone to bugs and security vulnerabilities (\eg based on the number of reported issues).
The presented quantitative analysis could be complemented by \changed{a full-fledged qualitative analysis based on in-depth interviews with package maintainers and users of these packages.} Such interviews can help to understand why package maintainers are reluctant to cross the 1.0.0 barrier, how they perceive \preone releases, and if they consider them different from \postone releases.

\section*{Acknowledgments}

This work was supported by the Fonds de la Recherche Scientifique -- FNRS under Grants number T.0017.18, O.0157.18F-RG43 and J.0151.20.

\bibliographystyle{elsarticle-num}
\bibliography{biblio}

\end{document}